\documentclass[aps,showpacs,prd,amsmath,twocolumn,amssymb,floatfix, 
superscriptaddress]{revtex4} 
 
\usepackage{epsfig} 
\usepackage{graphicx,psfrag} 
\usepackage{dcolumn}
\usepackage{bm}  
  
\newcommand{\beq}{\begin{equation}} 
\newcommand{\eeq}{\end{equation}}   
\newcommand{\bea}{\begin{eqnarray}} 
\newcommand{\eea}{\end{eqnarray}} 
\newcommand{\hf} {\frac{1}{2}}

\newcommand\eqn[1]     {Eq.\,(\ref{#1})}

\newcommand\fig[1]     {Fig.\,{\ref{#1}}}

\def\tu{{\tilde u}} 
\def\tg{{\tilde g}}
\def\tV{{\tilde V}} 
\def\tM{{\tilde M}}

\def\ord#1{{\cal O}(#1)} 
 
\def\mr#1{{\mathrm{#1}}} 
 
\def\eq#1{(\ref{#1})} 
 
\begin{document} 
 
\title{Comparison of renormalization group schemes for sine-Gordon type models}

\author{I. N\'andori} 
\affiliation{Institute of Nuclear Research of the Hungarian Academy of  
Sciences,H-4001 Debrecen, P.O.Box 51, Hungary} 
 
\author{S. Nagy} 
\affiliation{Department of Theoretical Physics, University of Debrecen, 
Debrecen, Hungary} 
 
\author{ K. Sailer} 
\affiliation{Department of Theoretical Physics, University of Debrecen, 
Debrecen, Hungary} 
 
\author{A. Trombettoni} 
\affiliation{International School for Advanced Studies,  
Via Beirut 2/4. I-34104, Trieste, Italy} 
 
\date{\today} 
 
\begin{abstract} 
The scheme-dependence of the renormalization group (RG) flow 
has been investigated in the local potential approximation for 
two-dimensional periodic, sine--Gordon type field-theoric models 
discussing the applicability of various functional RG methods in 
detail. It was shown that scheme-independent determination of 
such physical parameters is possible as the critical frequency 
(temperature) at which Kosterlitz-Thouless-Berezinskii type phase 
transition takes place in the sine-Gordon and the layered 
sine-Gordon models, and the critical ratio characterizing the 
Ising type phase transition of the massive sine-Gordon model.
For the latter case the Maxwell construction represents a strong 
constraint on the RG flow which results in a scheme-independent 
infrared value for the critical ratio. For the massive sine--Gordon 
model also the shrinking of the domain of the phase with spontaneously 
broken periodicity is shown to take place due to the quantum 
fluctuations. 
\end{abstract} 
 
 
\maketitle

\section{Introduction} 
Since the invention of the renormalization group (RG) method \cite{Wi1971}, 
it has been the main goal of functional RG to describe systems where the 
usual approximations (e.g. perturbation theory) failed. Strongly correlated 
electrons, critical phenomena, quark confinement are examples where the 
non-perturbative treatment is required. The RG equations are functional 
integro-differential equations, consequently, they can only be handled 
by using truncations (for reviews see \cite{rg}). The truncated RG flow 
depends on the particular choice of the so-called regulator function of 
the RG method, i.e. on the renormalization scheme, hence, the predicting 
power of the RG method is weakened. Indeed, once approximations are used a
dependence of physical results on the choice of the regulator function
is observed, consequently, any RG scheme makes sense if this dependence 
is weak. Therefore, it is of relevance to clarify how far the results 
obtained are independent (or at least weakly dependent) of the particular 
choice of the renormalization scheme used. 

The purpose of this paper is to investigate the RG scheme-dependence 
of two-dimensional (2D) periodic scalar field theories with possible 
inclusion of explicit symmetry breaking mass terms, i.e. the sine-Gordon 
(SG) \cite{sg_model}, the massive sine--Gordon (MSG) \cite{msg_model} 
and the multi-component layered sine--Gordon (LSG) \cite{lsg_model} 
models. Our motivation is twofold: (i) from the {\em methodical} point 
of view, 2D SG-type models represent a new platform to consider RG 
scheme-dependence since the comparison of results obtained by 
various RG schemes has been investigated previously for the O(N) 
symmetric polynomial scalar field theory mostly in three-dimensions
\cite{Li2001,scheme,LiPoSt2000,litimpolch,Mo2005} (ii) from the 
{\em phenomenological} point of view, 2D SG-type models have important 
direct realizations in high-energy and condensed matter physics, 
consequently, their physical parameters should be determined 
independently of the particular choice of the RG scheme used.

In particular, our goal is to consider under which conditions it is 
possible to determine scheme-independently three physical parameters, 
the critical frequency $\beta^2_c$ of the SG theory, the critical 
ratio $u/M^2$ of the MSG model and the layer-dependent critical 
value $\beta^2_c(N_L)$ of the LSG model. The SG scalar field theory 
(the Bose form of the massive Thirring fermionic model) represents 
the simplest non-trivial quantum field theory which is used to study 
the confinement mechanism. It has two phases separated by the critical 
value of the frequency $\beta_c^2 = 8\pi$. The 2D SG model belongs to 
the universality class of the 2D Coulomb gas and the 2D--XY spin model 
which have been applied to describe the Kosterlitz--Thouless--Berezinskii 
(KTB) \cite{KTB} phase transition of many condensed matter systems. Due 
to the exact mapping of the generating functional of the SG model onto 
the partition function of the Coulomb gas, $\beta_c^2$ is related to the 
critical temperature of the condensed matter system \cite{sg2,sg3} which 
can be measured directly. Moreover, the MSG model describes the vortex 
dynamics of 2D superconducting thin films \cite{PiVa2000,NaEtAl2007} 
and it is the bosonized version of the one-flavor massive Schwinger 
model, i.e. the 2D quantum electrodynamics (QED$_2$) 
\cite{msg_model,NaNaSaJe2005,NaEtAl2007msg,NaQED2} which possesses 
confinement properties. The Ising type phase transition of the MSG model 
is controlled by the dimensionless quantity $u/M^2$ where $u$ is the 
Fourier amplitude and $M$ is the scalar mass. It is related to the critical 
ratio of QED$_2$ which separates the confining and the half-asymptotic 
phases of the fermionic model which has been calculated by semiclassical 
and lattice methods \cite{BySrBuHa2002,HaKoCrMa1982,ScRa1983}. The LSG 
model is used to describe the vortex dynamics of magnetically coupled 
layered superconductors \cite{NaEtAl2007,Na2008,NaSa2006,Na2006,JeNaZJ2006} 
and considered to be the bosonized form of the multi-flavor Schwinger 
model (i.e. multiflavor QED$_2$) \cite{Na2008,NaNaSaJe2005,NaQED2_multi}. 
It is known that the critical frequency of the LSG model (and consequently 
the critical temperature of the corresponding condensed matter system) 
depends on the number of layers $N_L$ in a particular way 
$\beta^2_c(N_L)=8\pi N_{L}/(N_{L}-1)$ \cite{NaEtAl2007}. This layer 
number dependence can be observed experimentally \cite{blatter}. Therefore, 
the critical frequency $\beta^2_c$ of the SG theory, the critical ratio 
$u/M^2$ of the MSG model and the layer-dependent critical value 
$\beta^2_c(N_L)$ of the LSG model are physical parameters, consequently, 
it is an important issue to consider under which conditions it is possible 
to determine them scheme independently which is the main goal of the 
present work.

The structure of our paper is the following. Some general remarks on RG 
scheme-dependence is given in Section \ref{general}. The applicability 
of various RG schemes to the 2D SG model and the scheme-dependence of 
the critical frequency $\beta^2_c$ is discussed in Section \ref{sectsg}. 
In Section \ref{sectmsg} we investigate the scheme-dependence of the 
critical ratio of the MSG model which characterizes the boundary of the 
phases of the model. In Section \ref{sectlsg} we consider the dependence 
of the critical frequency $\beta^2_c(N_L)$ of the LSG model on the RG 
scheme used. Section \ref{sum} presents the summary and our concluding 
remarks. A brief overview of the frequently used RG methods is given 
in Appendix \ref{rgm} in order to remind the reader on their advantages 
and drawbacks and settling the notations used throughout the paper. 
Finally, in Appendix \ref{polyn} the scheme-dependence of the 
renormalization of the one-component polynomial scalar field theory 
is reviewed.

\section{Motivation and general remarks on scheme-dependence} 
\label{general}
In the framework of the Kadanoff-Wilson \cite{Wi1971} RG approach the 
differential RG transformations are realized via a blocking construction, 
the successive elimination of the degrees of freedom which lie above the 
running ultraviolet (UV) momentum cutoff $k$. Consequently, the effective 
theory defined by the (bare or effective) blocked action contains quantum 
fluctuations whose frequencies are smaller than the momentum cutoff. 
This procedure generates, e.g. the functional RG flow equation, 
\beq
k \partial_k \Gamma_k [\phi] = \hf \mathrm{Tr}  
\left( \Gamma^{(2)}_k [\phi] + R_k \right)^{-1}  
k \partial_k R_k 
\nonumber
\eeq 
for the effective action $\Gamma_k [\phi]$ when various types of regulator 
functions $R_k$ are used, where $\Gamma^{(2)}_k [\phi]$ denotes the second 
functional derivative of the effective action (see e.g. \cite{rg}). 
Here $R_k$ is a properly chosen infrared (IR) regulator function which 
fulfills a few basic constraints to ensure that $\Gamma_k$ approaches the 
bare action in the UV limit ($k\to\Lambda$) and the full quantum effective 
action in the IR limit ($k\to 0$). Indeed, various renormalization schemes 
are constructed in such a manner that the RG flow starts at the bare action 
and provides the effective action in the IR limit, so that the physical 
predictions (e.g. fixed points, critical exponents) are independent of the 
renormalization scheme particularly used. 

Since RG equations are functional partial differential equations it is not 
possible to solve them in general, hence, approximations are required. 
One of the commonly used systematic approximation is the truncated 
derivative expansion where the (bare or effective) action is expanded in 
powers of the derivative of the field,
\beq
\Gamma_k [\phi] = \int_x \left[V_k(\phi) 
+ Z_k(\phi) \hf (\partial_{\mu} \phi)^2 + ... \right].  
\nonumber
\eeq 
In the local potential approximation (LPA) higher derivative terms are 
neglected and the wave-function renormalization is set equal to constant, 
i.e. $Z_k \equiv 1$. The solution of the RG equations sometimes requires 
further approximations, e.g. the potential can be expanded in powers of the 
field variable (with a truncation at the power $N$)
\bea  
V_{k}(\phi) = \sum_{n=1}^N \, c_{n}(k) \, \phi^{n} 
\nonumber
\eea  
where the scale-dependence is encoded in the coupling constants $c_{n}(k)$.
Since the approximated RG flow depends on the choice of the regulator 
function the physical results could become scheme-dependent.

A general issue is the comparison of results obtained by various RG 
schemes \cite{Li2001,scheme,LiPoSt2000,litimpolch,Mo2005} which has been 
investigated for the O(N) symmetric polynomial scalar field theory in 
three dimensions in great detail. Some of the main results are as follows:
(i) for a given RG equation the explicit dependence on the regulator 
function disappears in the LPA, i.e. the results should not depend on 
the form of the regulator function if no further approximations (e.g. 
truncation in powers of the field) are used,
(ii) RG equations linearized around the trivial UV Gaussian fixed point 
($V_{*}=0$) provide the same UV scaling laws in various RG schemes, 
consequently, critical exponents of the Gaussian fixed point are 
scheme-independent in the LPA, 
(iii) non-trivial fixed points (like the Wilson-Fisher fixed point) and 
the critical exponents characterizing the scaling in their neighbourhood
obtained in the LPA become scheme-dependent, i.e. the estimates of these 
physical quantities depend on the particular choice of the RG scheme. 
Therefore, it is generally assumed that once approximations are used 
non-trivial fixed points and their critical behavior become scheme-dependent. 
However, in this paper we show that one can find models which have important 
physical realizations and their physical parameters (fixed points, critical 
ratio) can be obtained scheme-independently even in the LPA.

Our goal here is to consider the scheme-dependence of the low-energy 
behavior of various RG methods by considering the renormalization of 
the 2D periodic scalar field theory, i.e., the 2D SG model \cite{sg_model} 
with possible inclusion of explicit symmetry breaking mass terms, i.e., 
the MSG \cite{msg_model} and the multi-component LSG model \cite{lsg_model}. 
The functional RG approaches are investigated in the LPA. Four types of 
RG methods are compared: the Wegner--Houghton (WH--RG) \cite{WeHo1973}, 
the Polchinski (P--RG) \cite{Po1984}, the Callan-Symanzik type (CS--RG) 
\cite{internal} and the effective average action (EAA--RG) 
\cite{RiWe1990,We1993,Mo1994} approaches. In the latter case we use 
two types of regulator functions: the optimized \cite{Li2001} and the 
power-like (quartic) \cite{Mo1994} regulators. The phase structure of 
various SG type models obtained at quantum and classical level are also 
compared. 

As a rule, scheme-dependence is expected even in the exact (not truncated)
RG flow at intermediate scales between the UV and IR scales since various 
schemes realize the elimination of quantum fluctuations in a different 
manner. However, the disappearance of scheme-dependence of the exact RG 
flow is expected in the deep IR limit if no approximations are involved. 
In the UV limit even the LPA of the RG flow is able to produce results 
which are independent of the particular choice of the renormalization 
scheme since various RG equations linearized at the trivial Gaussian 
fixed point are the same in the LPA. Nevertheless, RG flows obtained in 
the LPA show spurious scheme-dependence at the non-trivial Wilson-Fisher 
fixed point. Therefore, it is a natural question to ask whether it is 
possible to obtain scheme-independent results at non-trivial fixed points. 
The answer can be affirmative if an additional constraint influences the 
RG flow. For example theories exhibiting spontaneous symmetry breaking 
have superuniversal effective IR behavior due to the Maxwell-cut in the 
symmetry broken phase. Superuniversality represents so strong constraints 
on the RG flow that the scheme-dependence disappears if the effects of the 
truncation of the functional subspace are under control by using a 
sufficiently large functional subspace. In this paper, we will show that 
this general view holds for the MSG model where scheme-independent results 
are obtained for the critical ratio $(u/M^2)_c$ being one of the typical 
IR characteristics of the Ising type phase transition of the MSG model. 
It will, however, also be shown that RG flow equations linearized at the 
UV Gaussian fixed point provide us scheme-dependent results for the 
critical ratio $(u/M^2)_c$ of the MSG model which characterizes the phase 
structure far beyond the validity of the UV scaling laws, in the deep IR 
region. 

Consequently, one expects that UV scaling laws cannot be used to determine
the critical behavior of any non-trivial fixed point situated in the IR or 
in the crossover regions. At cross-over scales between the UV and IR scaling 
regions even ``exact'' RG equations (where LPA is the only approximation 
used) produce results being influenced by the choice of renormalization 
scheme. Nevertheless, we shall show that the critical frequency 
$\beta^2_c =8\pi$ at which the SG model and the layer-dependent critical 
value $\beta^2_c(N_L) = 8\pi N_{L}/(N_{L}-1)$ at which the multi-component 
($N_{L}>1$) LSG models undergo a KTB type phase transition can be obtained 
exactly (independently of the invented RG scheme) even by the UV linearized 
RG flow. This rather surprising result is due to the following circumstances:  
(i) although UV linearized flow equations obtained in various RG schemes 
are different for the LSG model (where the linearization is performed in the 
periodic piece of the potential and not in the full potential), they provide 
us the same scheme-independent critical frequency, (ii) the KTB type phase 
transition of SG and LSG models is governed by the fundamental Fourier mode 
which has the same scaling law in the UV and IR regions, (iii) higher 
harmonics are generated by RG transformations and they have different UV 
and IR scalings but they do not influence the critical behavior.

\section{Sine--Gordon model} 
\label{sectsg}
In this section we compare the applicability of the frequently used 
functional RG methods to two-dimensional one-component Euclidean scalar 
fields with periodic self-interaction. In Appendix \ref{rgm} we review 
the notation and the main properties of various RG schemes used by us 
which are the following: the Wegner--Houghton (WH--RG), the Polchinski 
(P--RG), the Callan-Symanzik type (CS--RG) and the effective average 
action (EAA--RG) approaches. In the latter case we use the optimized 
and the quartic regulator functions. The WH--RG, the functional CS--RG, 
and the EAA--RG with power-like regulator with $b=1$ are equivalent 
methods in the LPA for $d=2$ even if spinodal instability (SI) \cite{tree}
occurs (see Appendix \ref{rgm}). Therefore, we shall compare the WH--RG, 
the EAA--RG with optimized and quartic regulators and the P--RG. It was 
shown \cite{Mo2005} that the P--RG and the EAA--RG with optimized 
regulator can be transformed into each other by a suitable Legendre 
transformation (in LPA), however, their singularity structures are 
different. We shall demonstrate this for the SG model since a SI, i.e. 
an IR singularity appears in the RG flow in the weak coupling phase 
but the P--RG method is not able to indicate this. Therefore, we shall 
conclude that any of the discussed RG schemes are expected to be 
applicable to the investigation of the IR behavior of the MSG model 
except the P--RG method which is inappropriate for quantitative analysis
in that case.
 
For the generalized SG model characterized by the local potential
\beq 
\label{periodic} 
\tilde V_{k}(\phi) = \sum_{n=1}^N \, \,  
{\tilde u}_{n}(k) \, \cos\left(n \, \beta \,\phi\right), 
\eeq 
exhibiting periodicity in the internal space the dimensionless couplings 
are represented by the Fourier amplitudes $\tilde u_{n}(k)$ and the 
`frequency' $\beta$ is a scale-independent, dimensionless parameter 
in the LPA. 
If the higher harmonics are neglected, i.e. for $N=1$ the generalized 
SG model  reduces to the well-known SG model \cite{sg_model}. We restrict 
ourselves to the RG analysis of bare models with $\tu_1(\Lambda)>0$ 
(positive fugacity for `charges' $\pm 1$ of the equivalent Coulomb gas). 
Another interesting generalization of the periodic model is the  
multi-component LSG model where the two-dimensional SG interaction 
terms are coupled by a particular mass matrix. This model 
receives important applications in high-energy and in low-temperature 
physics and its KTB-type phase transition has been discussed in 
\cite{NaEtAl2007,NaNaSaJe2005,NaSa2006,Na2006,JeNaZJ2006,Na2008}.
In Section \ref{sectlsg} we consider the scheme-dependence of the 
critical frequency of the multi-component LSG model.
 
\begin{figure}[ht] 
\includegraphics[width=8cm]{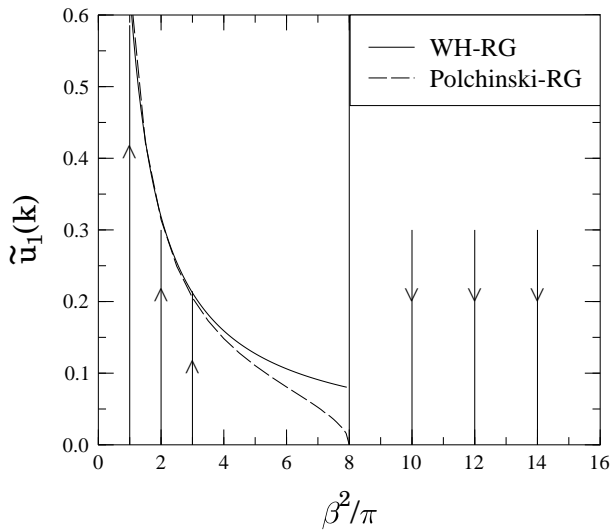} 
\caption{The phase structure of the SG model obtained in LPA
($\beta^2$ is scale-independent). The arrows indicate the direction of 
the RG flow. The non-trivial IR fixed points for $\beta^2<8\pi$ are 
obtained by the WH--RG (full line) and by the P--RG (dashed 
line) methods. The line of IR fixed points is given by the analytic 
expression $\tilde u_1(0) = 2/\beta^2$ for the WH--RG method for 
$\beta^2<8\pi$ and $\tilde u_1(0) \sim 0.03(8\pi-\beta^2)^{1/2}$ for 
the P--RG approach for $\beta^2\stackrel{<}{\sim} 8\pi$.
\label{sg_phases}} 
\end{figure}

Let us first briefly summarize the results of the WH--RG analysis 
of the generalized SG model obtained previously in the LPA 
\cite{sg2,sg3,NaEtAl2007sg,NaSaPo2006}. The phase structure (in LPA) 
is sketched in \fig{sg_phases} and has also been discussed in 
\cite{NaEtAl2007sg}.  
The $\beta^2$ axis represents the line of Gaussian fixed points, which 
are IR ones in the strong-coupling phase with $\beta^2>8\pi$ and UV 
ones in the weak-coupling phase with $\beta^2<8\pi$.  The two phases 
are separated by the vertical line at $\beta^2=8\pi$ where the KTB 
type \cite{KTB} phase transition takes place. 
In the figure only the single parameter axis $\tu_1$ is depicted. In 
order to get a more reliable picture of the phase diagram, one has to 
imagine a continuous sequence of non-intersecting infinite-dimensional 
hypersurfaces which are intersected by the $\beta^2$ axis at $\tu_n=0$, 
as well as, by the line of the IR fixed points in the weak-coupling phase 
at some $\tu_n(0)\not=0$. Then the hypersurface at  $\beta^2=8\pi$ separates 
the strong- and weak-coupling phases. In the UV limit ($k\leq \Lambda$) the 
linearized RG equations provide the UV scaling laws
\beq 
\label{uv_scaling} 
\tu_n(k) = \tu_n(\Lambda) 
\left(\frac{k}\Lambda\right)^{\frac{n^2\beta^2}{4\pi} -2} .
\eeq 
In the strong-coupling phase $(\beta^2>8\pi)$ all the Fourier amplitudes 
are UV irrelevant whereas in the weak-coupling one $(\beta^2<8\pi)$ the 
first few Fourier amplitudes become relevant, depending on the value of 
$\beta^2$. In both the strong- and weak-coupling phases the dimensionful 
blocked potential tends to a constant effective potential for $k\to 0$, 
being the only function which is simultaneously convex and periodic, but 
there is a significant difference in the IR scaling of the dimensionless 
couplings. The insertion of the ansatz \eq{periodic} into the WH-RG equation 
\eqn{derwh} yields the RG flow equations    
\beq\label{whtu} 
(2+k\partial_k)n\tu_n = \frac{\beta^2}{4\pi}n^3\tu_n  
+ \frac{\beta^2}{2}  
\sum_{s=1}^\infty s A_{n,s} (2+k\partial_k) \tu_s, 
\eeq 
for the couplings $\tu_n$ where 
$A_{n,s}(k)=(n-s)^2\tu_{|n-s|}-(n+s)^2\tu_{n+s}$. \eqn{derwh}, 
more precisely \eqn{WHdim} are valid unless SI arises. In the 
strong-coupling phase $\beta^2 > 8\pi$ no SI occurs, \eqn{whtu} 
holds at any scale and every Fourier amplitude is irrelevant. The IR 
scaling laws are given by 
\beq 
\label{ansatz} 
\tu_n(k) = c_n \left(\frac{k}\Lambda\right)^{n\eta} 
\eeq 
with $\eta = \beta^2/4\pi-2>0$ and the constants $c_n$ depending on 
$\beta^2$ and the single bare parameter $c_1=\tu_1(\Lambda)$ via the
recursion relation (for $n>1$) \cite{NaEtAl2007sg}
\beq 
c_n= 
\frac{\hf\beta^2 \sum_{s=1}^{n-1} (2+s\eta)  
s (n-s)^2 c_{n-s}c_s}{n(2+n\eta-n^2\frac{\beta^2}{4\pi})}. 
\eeq 
Here the well-justified approximation 
$A_{n,s}(k)\approx (n-s)^2\tu_{|n-s|}$ has been 
used. The dimensionless blocked potential becomes flat, i.e. 
all couplings $\tu_n$ vanish in the IR limit $k\to 0$. For 
$\beta^2<8\pi$ the SI occurs in the RG flow when the propagator 
diverges, $k_\mr{SI}^2+V''_{k_\mr{SI}}(\phi)=0$. The scaling laws  
just above the scale $k_\mr{SI}$ of the SI are given by \eq{ansatz},  
but now with $\eta<0$ which modifies the scaling essentially. Namely,  
all the Fourier amplitudes $\tu_n$ become relevant at the scale $k_{\mr{SI}}$.
Even more radical change of the IR scaling laws has been observed.
Making use of the tree-level blocking relation \eq{treedim} one finds 
that the SI results in the building up of a condensate of increasing 
amplitude $\rho$ for decreasing scale $k<k_{\mr{SI}}$. In the weak coupling 
phase the dimensionless effective potential remains a non-vanishing 
periodic one (the line of non-trivial IR fixed points in \fig{sg_phases}), 
graphically obtained by setting forth along the $\phi$ axis the section 
of the parabola
\begin{align}
\label{effpotmol} 
\tilde V_{k\to 0}(\phi)  = 
\frac2{\beta^2} 
\sum_{n=1}^\infty \frac{(-1)^{n+1}}{n^2} \cos(n\beta\phi)
= -\hf\phi^2   
\end{align}
with $\phi\in [-\pi/\beta,\pi/\beta]$ periodically. Each parabola section 
is the one of just the same parabola \eq{treewhpot}, one would find as 
the non-trivial fixed point of the polynomial theory. The periodic 
dimensionless effective potential is the continuous, sectionally 
differentiable, periodic solution of \eqn{sieq}. Let us note that 
\eqn{whtu} yields the fixed point equation
\beq\label{static} 
2n\tu_n = \frac{\beta^2}{4\pi}n^3\tu_n+\hf\beta^2 
\sum_{s=1}^\infty s A_{n,s} 2 \tu_s
\eeq 
if no SI occurs in the RG flow.

\begin{figure}[ht] 
\includegraphics[width=8cm]{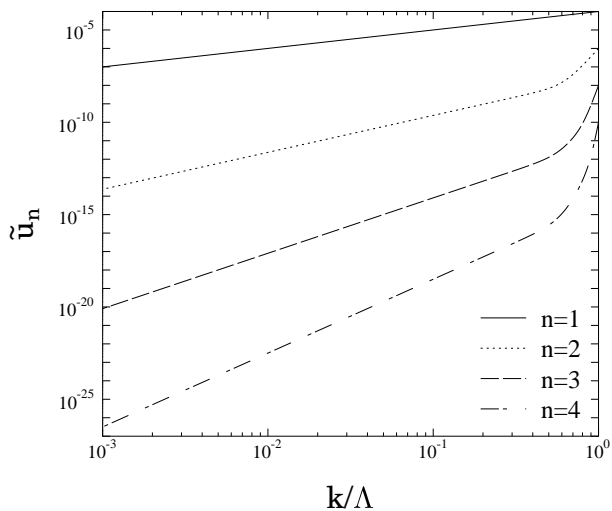} 
\caption{The P--RG flow of the first few dimensionless  
couplings of the generalized SG model for $\beta^2 = 12\pi$. 
\label{12pi}} 
\end{figure}

Using the same machinery in the framework of the P-RG, one obtains 
from \eqn{derpolch2} the flow equations 
\beq 
\label{polchtu} 
(2+k\partial_k)n\tu_n  
= \frac{\beta^2}{4\pi}n^3\tu_n - \hf\beta^2 
\sum_{s=1}^\infty s A_{n,s} 2 \tu_s . 
\eeq 
Inserting the ansatz \eq{ansatz} into \eqn{polchtu} we get 
\beq 
(2+n\eta)nc_n k^{n\eta}  
= \frac{\beta^2}{4\pi}n^3c_nk^{n\eta} - \frac{\beta^2}{2} 
\sum_{s=1}^\infty s A_{n,s} 2 c_s k^{s\eta}
\nonumber 
\eeq 
which gives $\eta = \beta^2/4\pi-2$ again. For $\beta^2>8\pi$ 
ansatz \eq{ansatz} works with the neglection of the term 
proportional to $\tu_{n+s}(k)$ in the expression of $A_{n,s}(k)$ 
similarly to the WH-RG and yields the recursion relation 
\beq 
c_n= - \frac{\hf\beta^2 \sum_{s=1}^{n-1}  2s(n-s)^2c_{n-s}c_s} 
{n(2+n\eta-n^2\frac{\beta^2}{4\pi})}, 
\label{c_n} 
\eeq 
with $c_1=\tu_1(\Lambda)$, 
$c_n = \tilde u_1^n(\Lambda) R_n$ for $n>1$, $R_1=1$ 
and the constants $R_n$ for $n>1$, satisfying the recursion 
relation \eq{c_n}, become independent of the bare couplings. 
In \fig{12pi} the flow of the first few couplings can be 
followed. All dimensionless couplings tend to zero in the IR 
limit $k\to 0$ again. 

\begin{figure}[ht] 
\includegraphics[width=8cm]{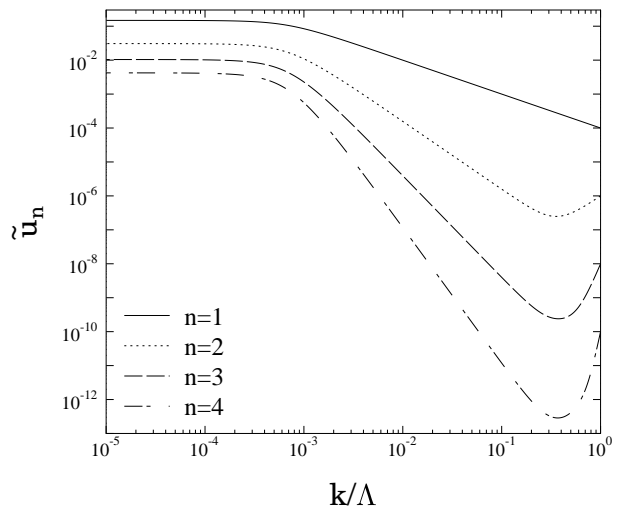} 
\caption{The flow of the first few dimensionless couplings of the  
SG model for $\beta^2=4\pi$ obtained by the P--RG method. 
\label{4pi}} 
\end{figure}

\begin{figure}[ht] 
\includegraphics[width=8cm]{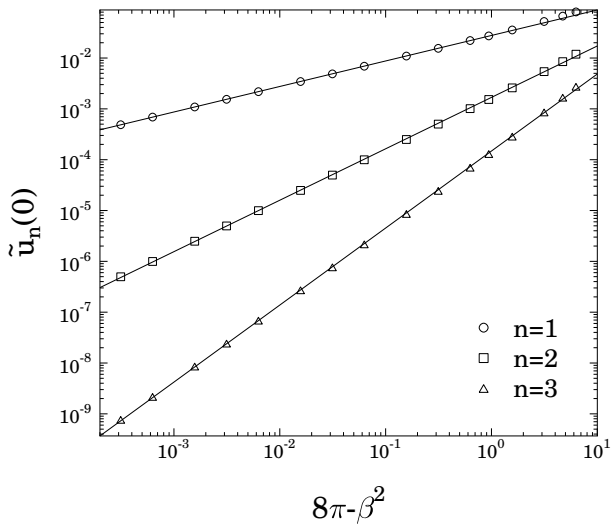} 
\caption{The IR values $\tu_n(0)$ of the dimensionless couplings  
show power law behavior in the vicinity of the critical value of  
$\beta^2$. The markers are obtained by solving the non-linear  
system of equations in \eqn{static}, and the lines are plotted 
by a fit giving $0.5,~1.0,~1.5$ for the slopes, respectively for 
$n=1,~2,~3$. 
\label{bscale}} 
\end{figure}

For $\beta^2<8\pi$ the numerical solution of the flow equations
\eq{polchtu} for the dimensionless couplings $\tu_n(k)$ follows 
the power law behavior of \eqn{ansatz}, now with $\eta<0$, in a 
wide range above a certain scale $k_c$, but the couplings $\tu_n(k)$ 
go to constant values in the deep IR region, as it is demonstrated 
in \fig{4pi}. Let us note that the inverse propagator vanishes around 
the scale $\sim k_c$ which signals the appearance of the SI (i.e. 
$k_c\sim k_{\rm{SI}}$). However, the P-RG flow equations do not 
exhibit any singularity at that scale $k_c$.
\eqn{polchtu} provides a scale independent, IR fixed point solution 
now satisfying \eqn{static}. 
Our numerical results show in \fig{bscale} for $\beta^2$ close to but 
below $8\pi$ ($\beta^2 \leq 8\pi$) that the $\beta$-dependence of 
$\tu_n(0)$ can be factored out as 
\beq\label{betafact} 
\tu_n(0) = (8\pi-\beta^2)^{n/2} a_n, 
\eeq 
where the numbers $a_n$ are independent of any parameter of 
the model. Therefore, the non-vanishing, universal dimensionless 
IR effective potential obtained by the P--RG method 
reminds one on the similar result of the WH--RG analysis, but the  
parabolic shape of the dimensionless effective potential in its 
periods cannot be recovered. As a consequence, the line of IR 
fixed points of the SG model in the weak coupling phase has been 
modified, see the dashed line in \fig{sg_phases}.

As to the ambiguity of the decision based on the numerics whether 
the SI does or does not occur during the flow, it is in order to make 
here an important remark. At the first glance one expects that the 
accuracy of such a decision will be increased by increasing the number 
of Fourier-modes, that of the couplings $\tu_n$ taken into account.
With decreasing scale $k$ the dimensionful periodic potential 
becomes rather flat in each period. According to our numerical 
experience it approaches rather smoothly the potential \eq{effpotmol} 
for $\beta^2\stackrel{<}{\sim} 4\pi$ and then the occurrence of the SI 
can be detected without any doubt. However, the matter of things becomes 
much worse for $0<8\pi-\beta^2\ll 8\pi$ when there occur numerical 
instabilities if one reconstructs the potential.
In such a case one cannot decide unambiguously with the numerical 
method based on the Fourier expansion of the potential whether the SI 
does occur indeed. Solving the RG equation derived in the LPA for the 
potential without using any further expansion seems to be a reliable 
way to settle this point, but it lies out of the scope of the present 
paper.

As to the next, let us turn now to the  discussion of the EAA--RG flow 
for the generalized SG model \eq{periodic}. Using the optimized regulator,
and deriving \eqn{litim_rg} with respect to the field and multiplying its 
both sides by $(1+\tV_k'')^2$, one obtains the evolution equations  
\begin{eqnarray}
\label{litimtu} 
&(2+k\partial_k) n \tu_n = \frac{\beta^2}{4\pi}n^3\tu_n  
+ \beta^2 \sum_{p=1}^\infty p A_{n,p} (2+k\partial_k)\tu_p 
\nonumber \\
&- \frac{1}{4}\beta^4 \sum_{q=1}^\infty \sum_{p=1}^\infty 
p A_{n,q} A_{q,p} (2+k\partial_k) \tu_p 
\end{eqnarray}
for the Fourier amplitudes $\tu_n$.
%
For $\beta^2>8\pi$ the IR scaling laws are given again by \eqn{ansatz} 
with $\eta = \beta^2/(4\pi) -2>0$ and the recursion relation for $c_n$'s  
\begin{eqnarray} 
c_n = \frac{\beta^2 \sum_{p=1}^{n-1} (2+p\eta) p
(n-p)^2 c_{n-p} c_{p}}{n(2+n\eta-n^2\frac{\beta^2}{4\pi})} 
\hskip 1cm
\\
-\frac{\frac{1}{4}\beta^4 \sum_{q=1}^{n-1}\sum_{p=1}^{q-1} 
(2+p\eta) p (n-q)^2 (q-p)^2 c_{n-q} c_{q-p} c_p}
{n(2+n\eta-n^2\frac{\beta^2}{4\pi})}.
\nonumber 
\end{eqnarray}
For $\beta^2<8\pi$ the numerical solution of the system \eq{litimtu} 
of the coupled flow equations exhibits the following features. If one 
takes into account a sufficiently large number of Fourier modes, one 
finds that there exists a non-vanishing scale $k_{\mr{SI}}$ at which the 
inverse propagator vanishes that signals the presence of the SI.
The numerical problem of deciding whether the SI occurs or does not,
remains just the same as in the WH-RG framework. For $\beta^2$  in the 
vicinity of the critical value $8\pi$ and restricting oneself to the 
first few Fourier amplitudes, one finds that the inverse propagator 
does not vanish in the IR region. In this case the IR effective 
potential is similar to that obtained by the P--RG, with the scaling 
property \eq{betafact} (but with different numbers $a_n$).

Let us now discuss the flow in the framework of the EAA--RG with 
quartic regulator. For $\beta^2>8\pi$ the SI does not occur. 
On the basis of the result of the WH-RG analysis, for $\beta^2<8\pi$ 
one would expect the occurring of the SI, i.e. vanishing of 
$1 + \hf\tilde V''_{k}(\phi)$ at some scale $k_{\mr{SI}}\not=0$. For the 
single mode potential $\tilde V_k(\phi)=\tu(k) \cos(\beta \phi)$ 
this happens for $\phi=0$ if $1 - \hf\tu(k_{\mr{SI}})\beta^2=0$ and then 
$1 - \hf\tilde V''_{k_{\mr{SI}}}(\pi/\beta)=0$ holds as well. Therefore, 
the momentum scale $k_{+}$ where the quartic regulator RG is 
non-analytic and the scale of the SI $k_{-}=k_{\rm{SI}}$ coincide 
for the periodic model. The decision based on numerics on the 
existence of the SI suffers from the same problems as for the other
RG schemes, and  the IR scaling laws are also expected to be similar 
to those obtained on the basis of Eqs. \eq{whtu}, \eq{polchtu}, and 
\eq{litimtu}.  Since the SI occurs when the propagator in the right 
hand side of \eqn{general_effective_rg} develops a pole, and 
periodicity should not be violated, the IR fixed point potential 
should be obtained by setting forth periodically the parabola section 
in $[-\pi/\beta,\pi/\beta]$ given by \eqn{cb} and \eqn{eaafppot}. 

In conclusion, the IR scaling laws determined by various RG 
methods are qualitatively the same in the strong coupling phase 
of the generalized SG model. In the framework of WH--RG and the 
EAA-RG with the power-law regulator, the SI has been treated 
explicitly in the weak coupling phase which provides a reliable 
determination of the IR scaling laws even in that phase. In this
case the functional form of the low-energy effective potential 
is found to be the same but its exact value is scheme-dependent
(see e.g. \eqn{cb} and \eqn{eaafppot}). However, the potential can 
always be rescaled by a constant which leaves the physical results 
unchanged, consequently, only the functional form of the IR 
effective potential is of physical significance.

\section{Massive sine-Gordon Model} 
\label{sectmsg}
The MSG model for the number of dimensions $d=2$, characterized by the 
dimensionless bare potential 
\begin{equation}
\label{baremsg}
{\tilde V}_{k=\Lambda}(\phi) = 
\hf {\tilde M}^2_{\Lambda} \phi^2 + {\tilde u}(\Lambda) \cos(\beta\phi)
\end{equation}
exhibits two phases. For $\beta^2>8\pi$ only the phase with explicitly 
broken periodicity is present, and that phase extends to $\beta^2<8\pi$ 
if the bare dimensionless ratio $\tu(\Lambda)/\tM_{\Lambda}^2$ is smaller 
than a critical upper bound $[\tu(\Lambda)/\tM^2_{\Lambda}]_c$ depending 
on the parameters $\beta^2$ and $\tM^2_{\Lambda}$. For $\beta^2<8\pi$ and 
$\tu(\Lambda)/\tM^2_{\Lambda} > [\tu(\Lambda)/\tM^2_{\Lambda}]_c$ the 
periodicity is spontaneously broken, and the RG trajectories in that phase 
merge into a single trajectory in the deep IR region which is characterized 
by the unique ratio $\tu(k\to 0)/\tM^2_{k\to 0}=[\tu/\tM^2]_c=[u/M^2]_c$ 
depending on $\beta^2$ only. Our main goal is to discuss whether the 
determination of that unique ratio does depend on the choice of the 
previously discussed renormalization schemes. Let us note that the 
$Z_2$ symmetry $\phi\to -\phi$ can be used to distinguish the phases of 
the MSG model, it suffers a spontaneous breakdown in the phase with 
spontaneously broken periodicity when a condensate appears.

The phase structure of the MSG model has been discussed in the framework 
of the WH-RG in the LPA in \cite{NaEtAl2007msg} making use of the more 
general ansatz
\bea 
\label{msg} 
\tilde V_{k}(\phi) =  
\hf {\tilde M}_k^2 \, \phi^2 +  
\sum_{n=1}^N {\tilde u_n}(k)\cos (n \, \beta \, \phi)  
\eea 
for the local potential. (The parameter $\tu$ in the expression 
\eq{baremsg} corresponds to $\tu_1(\Lambda)$ and we are looking for the 
unique value of $\tu_1(k\to 0)/M^2_{k\to 0}=[u/M^2]_c$ in the deep IR 
region for the phase with spontaneously broken periodicity.) The periodic 
piece of the potential \eq{msg} possesses the discrete symmetry under the 
shift $\phi \to \phi + 2\pi/\beta$ of the field variable. This is the 
symmetry the generalized SG model with the potential \eq{periodic} exhibits, 
and -- as discussed in the previous section --  in the weak coupling phase 
($\beta^2 < 8\pi$) of the generalized SG model a condensate appears in 
the IR limit, due to which the periodicity is broken spontaneously.
Due to the explicit mass term the MSG model offers the opportunity to 
investigate the interplay between the spontaneous and explicit breaking 
of periodicity. The following features of the phase structure have been 
determined in the framework of the WH-RG method in the LPA in 
\cite{NaEtAl2007msg,NaSa2006}:
\begin{enumerate}
\item The dimensionful mass $M^2$ remains constant during the blocking 
which provides the trivial scaling $\tilde M^2_k = k^{-2} M^2$ for the
dimensionless mass.  In the LPA (when no wave-function renormalization 
is incorporated) also the parameter $\beta^2$ is constant during the flow.
\item The linearized WH--RG flow equation
\begin{equation} 
(2+ k\partial_k ){\tilde u_1}(k) = \frac{\beta^2}{4\pi}  
{\tilde u_1}(k) \frac{k^2}{k^2 + M^2} 
\end{equation} 
for the coupling $\tu_1(k)$ of the fundamental mode of the local potential
provides the UV scaling law
\begin{equation}
\label{msguv} 
{\tilde u_1}(k) =  {\tilde u_1}(\Lambda)  
\left(\frac{k}{\Lambda}\right)^{-2} 
\left(\frac{k^2 + M^2} 
{\Lambda^2 + M^2}\right)^{\frac{\beta^2}{8\pi}} .
\end{equation} 
The coupling $\tilde u_1(k)$ is relevant in the UV scaling region,
so that the scaling law \eq{msguv} looses its validity for IR scales,
irrespectively of $\beta^2$. Nevertheless, it was observed numerically 
that \eqn{msguv} provides a rather good description of the RG flow for 
scales $k\stackrel{>}{\sim}M$ for $\beta^2>8\pi$. The critical ratio
\begin{equation}
\label{barecrit}
\left\lbrack\frac{\tu_1(\Lambda)}{\tM^2_{\Lambda}}\right\rbrack_c 
= \frac{2}{\beta^2} 
\left(\frac{1+M^2_{\Lambda}}{2 M^2_{\Lambda}}\right)^{\frac{\beta^2}{8\pi}} 
\end{equation}
for the bare parameters of the MSG model has also been discussed in
the framework of the linearized RG (see Eq.(9) of \cite{NaEtAl2007msg}).
\item Inserting the ansatz \eq{msg} into the WH-RG equation 
\eqn{derwh} yields the RG flow equations    
\begin{eqnarray}
\label{msg_whtu} 
&&(1+ \frac{M^2}{k^2})(2+k\partial_k)n\tu_n = 
\frac{\beta^2}{4\pi}n^3\tu_n +
\\
&&+ \frac{\beta^2}{2}  
\sum_{s=1}^\infty s A_{n,s} (2+k\partial_k) \tu_s, 
\nonumber
\end{eqnarray}
for the couplings $\tu_n(k)$. 
For $k^2\gg M^2$ the RG flow is close to that of the SG model.
For strong coupling $\beta^2>8\pi$ no SI occurs during the flow, the 
inequality $1+\tV_k''>0$ holds for all scales $k$. The numerical solution 
of Eq. \eq{msg_whtu} provides scaling laws for $M<k\ll\Lambda$ not 
differing significantly from those of the SG model, and below the scale 
$k\sim \ord{M}$ the trivial scaling $\tu_n(k)\sim k^{-2}$ occurs. The 
parameter region with $\beta^2>8\pi$ belongs to the phase with explicit 
breaking of  periodicity.
\item For weak coupling $\beta^2<8\pi$ two phases appear.

On the one hand, the RG flow on the trajectories started at $\beta^2<8\pi$ 
and $\tu_1(\Lambda)/M^2_{\Lambda} > [u_1(\Lambda)/M^2_{\Lambda}]_c$
develops SI at some finite scale $k_{\mr{SI}}>M$, so that one has to evaluate 
the flow using the tree-level blocking relation \eq{treedim}. In this case 
the parameters $\tu_n(k)$ become superuniversal (i.e. independent of the 
bare parameters) for $k<k_{\mr{SI}}$ signaling the presence of the condensate 
and the spontaneous breaking of periodicity, similarly to the behaviour of 
the weak coupling phase of the SG model. It should be noticed that the 
detection of occurring SI in the evaluation of the trajectories suffers 
the same ambiguity for $\beta^2$ close to but smaller than $8\pi$ as in 
the case of the generalized SG model. For $k<k_{\mr{SI}}$ the tree-level 
evolution yields the periodic piece of the potential with parabola sections, 
\begin{eqnarray}
\label{msgpar} 
U_{k}(\phi)&=& 
(k^2 + M^2) \frac{2}{\beta^2} 
\sum_{n=1}^\infty \frac{(-1)^{n+1}}{n^2}
\cos(n\beta\phi) 
\nonumber \\ 
&=& -\hf(k^2 + M^2) \phi^2 
\end{eqnarray}
for $\phi\in [-\pi/\beta,\pi/\beta]$. Consequently, the first dimensionful 
Fourier amplitude tends to the constant $u_1(k\to 0) =\frac{2}{\beta^2} M^2$,
i.e. the IR value of the critical ratio is $[u/M^2]_c=2/\beta^2$.  
The parabolic potential \eq{msgpar} represents the non-trivial solution of 
the differential equation $k^2+V''_k(\phi)=k^2 +M^2 +U''_k(\phi)=0$. The 
effective potential $V_{k\to 0}=U_{k\to0}+\hf M^2\phi^2=-\hf k^2\phi^2$ is 
superuniversal, i.e. independent of the bare parameters $\tu_n(\Lambda)$.

On the other hand, the phase with explicit breaking of periodicity 
extends to the region with $\beta^2<8\pi$ bounded from above by
$[u_1(\Lambda)/M^2_{\Lambda}]_c$. In this phase no SI occurs along the 
RG trajectories and below the scale $k=M$ the Fourier amplitudes 
start to scale as $\tu_n(k)\sim k^{-2}$. In this case the effective 
potential $U_{k\to 0}$ turns out to depend on the single bare parameter 
$\tu_1(\Lambda)$.
\end{enumerate}

\begin{figure}[ht]
\includegraphics[width=8cm]{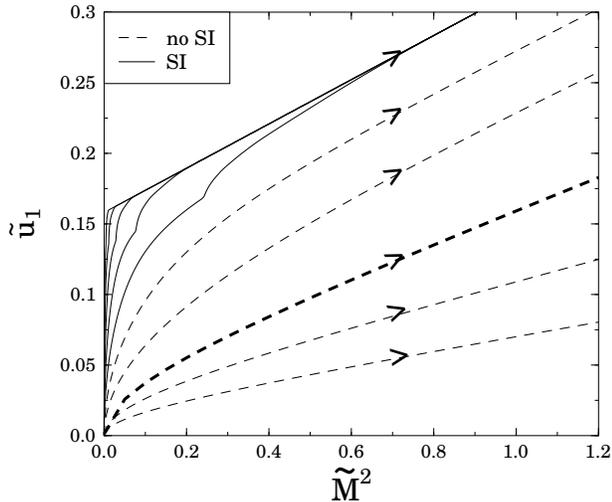}
\caption{Phase structure of the MSG model in LPA for $\beta^2=4\pi$
obtained by the full WH--RG method. The arrows indicate the direction 
of the flow. The numerical solution of the exact RG equation shows  
that the separatrix obtained by the linearized flow \eq{rgcrit2} 
(wide dashed line) is modified by higher order corrections.
\label{msg_phases}}
\end{figure}

The phase diagram of the MSG model for $\beta^2=4\pi$ is depicted in 
\fig{msg_phases}
where one can see that the RG trajectories in the phase 
with spontaneously broken periodicity merge into a single trajectory in 
the deep IR limit (full lines). It is worthwhile noticing that the KTB 
phase transition exhibited by the massless SG model disappears in the 
MSG model due to the presence of the explicit mass term, as demonstrated 
in \cite{NaEtAl2007msg,NaSa2006} using WH--RG and CS--RG methods.

\begin{figure}[ht] 
\includegraphics[width=8cm]{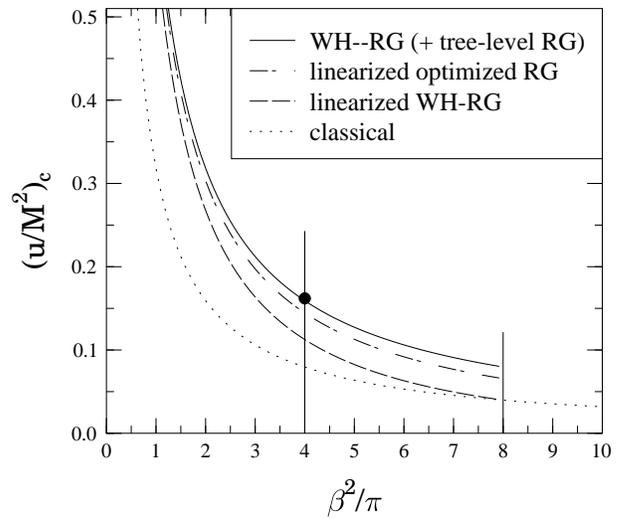} 
\caption{The critical ratio which separates the phases of the MSG 
model obtained at classical level $(u/M^2)_c = 1/\beta^2$ is compared 
to that of determined by WH-RG and EAA--RG methods. The full circle 
represents the results $(u/M^2)_c = 0.156-0.168$ calculated by 
lattice methods \cite{BySrBuHa2002,HaKoCrMa1982,ScRa1983} at 
$\beta^2=4\pi$. The solid line stands for $(u/M^2)_c = 2/\beta^2$ 
which is obtained by the full WH--RG (+ tree-level RG) method. (The 
full EAA--RG (+ tree-level RG) method provides the same critical 
ratio (solid line). 
The dashed line is determined by the linearized WH--RG 
equation and it is given by \eq{lincritwh}. The dashed-dotted line 
is obtained by the linearized optimized EAA--RG approach, see 
Eq.\eq{lincritlitim} which gives better result then the linearized 
WH--RG. For $\beta^2=4\pi$ it is shown that only the usage of the 
full RG equation gives reliable result for the critical ratio i.e. 
$(u/M^2)_c = 0.159$ which coincides to the results of the lattice 
calculations.
\label{critical}} 
\end{figure}

Let us consider the critical RG trajectory on the $\tu_1 - \tM^2$ plane
which separates the phases of the MSG model. At linearized level this is 
constructed from \eqn{msguv} as
\begin{equation}
\label{rgcrit}
{\tilde u_{1c}}(\tM^2) =  {\tilde u_{1c}}(\Lambda)  
\left(\frac{\tM^2}{\tM^2_{\Lambda}}\right)^{1-\frac{\beta^2}{8\pi}} 
\left(\frac{1 + \tM^2}{1+ \tM^2_{\Lambda}}\right)^{\frac{\beta^2}{8\pi}} 
\end{equation}
where $\tM^2 = \tM_k^2 = k^{-2} M^2$ is the running coupling and 
$\tM_{\Lambda}^2 = \Lambda^{-2} M^2$ is constant. By inserting the 
critical ratio \eq{barecrit} obtained for the initial bare values into 
\eqn{rgcrit} one finds
\begin{equation}
\label{rgcrit2}
{\tilde u_{1c}}(\tM^2) =  \frac{2}{\beta^2} \tM^2 
\left(\frac{1 + \tM^2}{2\tM^2}\right)^{\frac{\beta^2}{8\pi}}. 
\end{equation}
At the mass-scale it gives $\tu_{1c}(\tM^2=1) = 2/\beta^2 \approx 0.159$
(see the wide dashed line in \fig{msg_phases}). From \eqn{rgcrit2} one 
can read off the IR value of the critical ratio 
\begin{equation}
\label{lincritwh}
\left\lbrack\frac{u}{M^2}\right\rbrack_c = \mr{lim}_{\tM^2 \to \infty} 
\left(\frac{\tu_{1c}}{\tM^2}\right) = 
\frac{2}{\beta^2}  \left(\frac{1}{2}\right)^{\frac{\beta^2}{8\pi}}
\end{equation}
which is smaller than the superuniversal ratio $[u/M^2]_c =2/\beta^2$
obtained by the exact WH--RG method (for $\beta^2\neq 0$) and coincides 
with it only in the limit $\beta^2\to 0$, see the dashed line in 
\fig{critical}. Indeed, the numerical solution of the exact WH--RG 
equation \eq{msg_whtu} shows that one finds RG trajectories (see, 
\fig{msg_phases}) which lie above the critical one obtained in the 
framework of the linearized WH--RG. 

In \fig{critical} we plotted the superuniversal ratio 
$[u/M^2]_c=u_1(k\to 0)/M^2$ against the parameter $\beta^2$ obtained 
in the framework of the WH--RG. According to our numerical results 
the effective potential \eq{msgpar} implies $[u/M^2]_c=2/\beta^2$. 
According to lattice results \cite{BySrBuHa2002} the phase transition 
of the MSG model is assumed to belong to the same universality class 
as the two dimensional Ising model (with the critical exponents 
$\nu= 1, \beta=1/8$ \footnote{Remind that the critical exponent 
$\beta$ has nothing to do with the `frequency' parameter $\beta$ of 
the MSG model.}). 
Since the MSG model for $\beta^2 = 4\pi$ represents the bosonized 
version of QED$_2$, the ratio $[u/M^2]_c = 1/(2\pi) \approx 0.1591$ 
obtained by the WH--RG method can also be used to determine the phase 
transition point of QED$_2$ given as $[m/g]_c  = 2\sqrt{\pi}e^{-\gamma}  
[u/M^2]_c \approx 0.3168$ where $m$ is the fermion mass, $g$ is the 
coupling between the fermionic and the gauge field and 
$\gamma \approx 0.5774$ is the Euler's constant.  It is shown in 
\fig{critical} that our WH--RG result is in good agreement with the 
result $[m/g]_c =$ $0.31$--$0.33$ obtained by density matrix and 
lattice methods \cite{BySrBuHa2002,HaKoCrMa1982,ScRa1983}.

\begin{figure}[ht] 
\includegraphics[width=8cm]{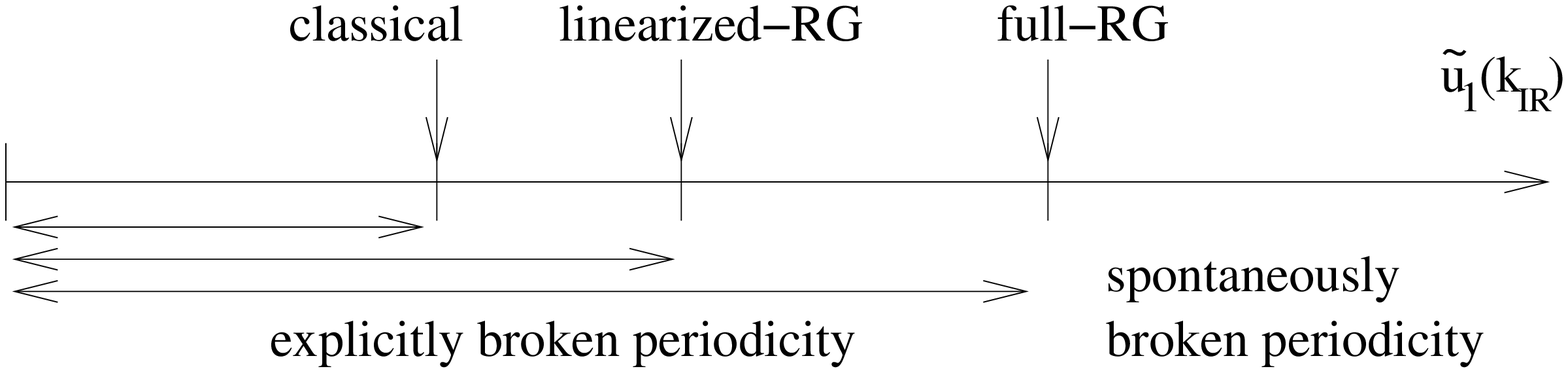} 
\caption{Regions of explicitly 
and spontaneously broken periodicity of the MSG model obtained by 
various approximations. The critical value of the fundamental 
Fourier amplitude $\tu_1(k_{\mr{IR}})$ is calculated by classical, 
linearized-RG and full-RG methods at some IR momentum scale 
$k_{\mr{IR}}<< M << \Lambda$. The effect of quantum fluctuations 
leads to the shrinking of the region of spontaneous breakdown of 
symmetry. 
\label{regions}} 
\end{figure}

It is illustrative to compare the critical ratio $[u/M^2]_c=2/\beta^2$
obtained by the WH--RG method with the estimation based on the classical 
potential \eq{baremsg} by solving $V_\Lambda''(\phi=0)=0$ which gives 
$[u/M^2]_c = 1/\beta^2$ i.e., 
$[m/g]_c  = 2\sqrt{\pi}e^{-\gamma} /\beta^2 \approx 0.1584$, a value 
being one-half of the one obtained by taking the quantum fluctuations 
into account. This means that the slope of the critical RG trajectory 
which separates the phases of the MSG model is steeper when the effect 
of the quantum fluctuations is taken into account, i.e. the latter 
enlarges the parameter region with explicitly broken symmetry, 
see \fig{regions}. 
In general, the effect of quantum fluctuations leads to the shrinking 
of the region of spontaneous breakdown of symmetry. For example, the 
polynomial model defined by \eqn{polynom} exhibits a double-well bare 
potential for the initial conditions with  $g_2(\Lambda)<0$ chosen in 
Appendix \ref{polyn}, but the running mass $g_2(k)$ changes its sign due 
to quantum effects (see \fig{fig_poly}) and the minimum of the effective  
potential remains at $\phi=0$.
Let us note that the classical analysis is not able to distinguish 
between the weak ($\beta^2<8\pi$) and the strong ($\beta^2>8\pi$) 
coupling regimes, it determines the critical ratio 
$[u/M^2]_c = 1/\beta^2$ independently of the actual choice of the 
frequency $\beta^2$. Only the detailed RG study of the MSG model can 
show that for $\beta^2>8\pi$ the SI, i.e. the 
condensate does not appear, hence the critical value cannot be 
extended for $\beta^2>8\pi$, see the vertical line in \fig{critical}. 

In Appendix \ref{polyn}, it was demonstrated that beyond the mass scale 
in the deep IR limit the RG flow of the polynomial scalar field theory 
(in the phase with unbroken $Z_2$ symmetry) determined by the P--RG 
differs of the usual one obtained by other RG methods. It was also shown 
in Section \ref{sectsg} that the P--RG approach is not able to signal 
the appearance of SI for the weak coupling phase of the SG model. 
Consequently, the P--RG is inappropriate to investigate the IR scaling 
of the MSG model quantitatively, hence, we do not discuss the RG flow
of the MSG model in the framework of the P--RG method. 

Let us turn to the discussion of the MSG model \eq{msg} in the 
framework of the EAA--RG. Considerations similar to those made 
in the case of the WH--RG lead to the linearized evolution equation   
\begin{equation} 
(2+ k\partial_k ){\tilde u_1} = \frac{\beta^2}{4\pi}  
{\tilde u_1} 
\frac{k^4}{(k^2 + M^2)^2} 
\end{equation} 
for the fundamental Fourier amplitude when the optimized 
regulator is made of use. Here $M^2$ and $\beta^2$ are again 
scale-independent parameters and the analytic solution, 
\begin{eqnarray}
\label{msguvop} 
{\tilde u_1}(k) =  {\tilde u_1}(\Lambda)  
\left(\frac{k}{\Lambda}\right)^{-2} 
\left(\frac{k^2 + M^2} 
{\Lambda^2 + M^2}\right)^{\frac{\beta^2}{8\pi}} 
\nonumber \\ 
\exp\left( \frac{\beta^2}{8\pi}\left[\frac{M^2}{k^2 + M^2}-  
\frac{M^2}{\Lambda^2 + M^2}\right]\right), 
\end{eqnarray} 
provides qualitatively the same UV scaling law  as that of 
\eqn{msguv} and indicates the absence of the KTB-type phase
transition. Let us determine the critical RG trajectory on the 
$\tu_1 - \tM^2$ plane at linearized level,
\begin{eqnarray}
\label{litimrgcrit}
{\tilde u_{1c}}(\tM^2) =  {\tilde u_{1c}}(\Lambda)  
\left(\frac{\tM^2}{\tM^2_{\Lambda}}\right)^{1-\frac{\beta^2}{8\pi}} 
\left(\frac{1 + \tM^2}{1+ \tM^2_{\Lambda}}\right)^{\frac{\beta^2}{8\pi}} 
\nonumber \\
\exp{\left[\frac{\beta^2}{8\pi} \left(\frac{\tM^2}{1+\tM^2} 
- \frac{\tM^2_{\Lambda}}{1+\tM^2_{\Lambda}}\right)\right]}.
\end{eqnarray}
The critical initial bare value $\tu_{1c}(\Lambda)$ is fixed by the 
condition $\tu_{1c}(\tM^2=1) = 2/\beta^2$ similarly to the WH--RG
case, then one finds 
\begin{eqnarray}
\label{litimrgcrit2}
{\tilde u_{1c}}(\tM^2) =  \frac{2}{\beta^2} \tM^2 
\left(\frac{1 + \tM^2}{2\tM^2}\right)^{\frac{\beta^2}{8\pi}}
\nonumber \\ 
\exp{\left[\frac{\beta^2}{8\pi} \left(\frac{\tM^2}{1+\tM^2} 
- \hf\right)\right]}.
\end{eqnarray}
From \eq{litimrgcrit2} one can read off the IR value of the critical 
ratio 
\begin{equation}
\label{lincritlitim}
\left\lbrack\frac{u}{M^2}\right\rbrack_c = \mr{lim}_{\tM^2 \to \infty} 
\frac{\tu_{1c}}{\tM^2} = 
\frac{2}{\beta^2}  \left(\frac{1}{2}\right)^{\frac{\beta^2}{8\pi}}
e^{\frac{\beta^2}{8\pi}}
\end{equation}
which is smaller then the superuniversal ratio $[u/M^2]_c =2/\beta^2$
obtained by the exact WH--RG method but gives a better result then 
Eq. \eq{lincritwh} obtained by the linearized WH--RG equation, 
see the dashed-dotted line in \fig{critical}. Indeed, the optimized
regulator has been constructed to achieve the best convergence of the 
truncated RG equation, consequently, between various RG equations 
considered at the same order of the truncation, the optimized one 
gives the closest result to the exact one \cite{Li2001}.

However, neither the linearized WH--RG nor the linearized optimized
EAA--RG enables one to map the phase structure of the MSG model in a 
reliable manner, i.e. to determine the exact critical ratio $[u/M^2]_c$ 
which can only be obtained by solving the full RG equation in the IR 
limit $k\to 0$. As shown in the framework of the WH--RG method, 
if SI arises, its appropriate treatment is necessary  to determine 
the effective potential beyond the scale of SI and the critical 
ratio $[u/M^2]_c$. The flow  equation \eq{litim_rg} has a pole at 
$U''_k +M^2= -k^2$, consequently, the IR effective potential beyond 
the scale $k_{\mr{SI}}$ is found to be identical with that of 
\eqn{treewhpot} obtained by the WH--RG method. Therefore, the 
critical ratio $[u/M^2]_c$ coincides with that determined by the 
WH--RG approach (see the solid line in \fig{critical}).

Finally, in the framework of the  EAA--RG with the quartic regulator 
the linearization of \eqn{morris_b2_rg} results in the flow equation 
\begin{eqnarray}
\label{qrli} 
(2+ k\partial_k ){\tilde u_1}&=& \frac{\beta^2}{4\pi}  
{\tilde u_1} 
\frac{k^4}{\frac{1}{4} M^4}\mu_k^4  \\  
&\times&\Biggl\{\begin{array}{lll} 
1+\mu_k^2({\mathrm{arctg}}\mu_k^2 -\frac{\pi}{2}),
&{\mbox{for}}& \hf M^2<k^2\cr 
-1-\frac{\mu_k^2}{2} \ln \left( \frac{\mu_k^2 -1}{\mu_k^2 +1} \right), 
&{\mbox{for}}& \hf M^2>k^2 
\end{array} 
\nonumber 
\end{eqnarray} 
with $\mu_k^2 = \hf M^2/\sqrt{\vert k^4-\frac{1}{4}M^4\vert}$ and 
the scale-independent parameters $M^2$ and $\beta^2$. Eq. \eq{qrli} 
provides again an UV scaling law which is very much like those 
of \eqn{msguv} and \eqn{msguvop}. In order to find the validity range 
of the mass-corrected UV scaling law and to map the phase structure, 
one has to solve \eqn{morris_b2_rg} numerically. For sufficiently 
large initial values of $\tu_1(\Lambda)$ an IR singularity appears in 
the RG flow at some scale $k_{\mr{SI}}$, similarly to what happened in the 
framework of the  WH--RG  and the EAA--RG with the optimized regulator. 
One has to use Eq.\eq{eaasi}, i.e. $M^2+U_{k}''=-C(b)k^2$ to treat the 
SI explicitly and arrives at $U_{k}=-\hf[C(b)k^2+M^2]\phi^2$ and the 
IR effective potential \eq{eaafppot} with $C(2)=2$, i.e. 
$V_{k\to 0} = - k^2  \phi^2$. Since $U_{k=0}=-\hf M^2\phi^2$ is 
independent of the scheme-dependent constant $C(b)$, the critical 
ratio turns out to be independent of the scheme as well. Further on 
the potential $U_{k=0}$ and, consequently the critical ratio 
$[u_1/M^2]_c = 2/\beta^2$ are just the same as those found by the 
WH--RG method.
 
One can conclude, on the one hand that the  WH--RG, the CS--RG (being 
equivalent now with the WH--RG), and the EAA--RG methods enable one 
to determine the same phase structure and critical ratio $[u/M^2]_c$ 
for the MSG model. On the other hand the P--RG method is found to be
inappropriate to follow the RG trajectories beyond the mass scale and 
to perform a quantitative analysis of the IR behavior of the MSG model. 
It was also shown that the truncated RG equations produce a 
scheme-dependent critical ratio but any exact RG (LPA is the only 
approximation used) gives the same (scheme-independent) result which 
coincides with the results of density matrix and lattice methods.

\section{Layered sine--Gordon model} 
\label{sectlsg}
In this section we show that the critical frequency which 
separates the phases of an SG-type model which undergoes a 
Kosterlitz-Thouless-Berezinskii type phase transition similar to 
the one for the two-dimensional SG model can be obtained exactly 
(and scheme-independently) by the linearized RG flow. As an example 
we consider the renormalization scheme-dependence of the LSG model, 
i.e. the multi-component scalar field theory where the two-dimensional 
periodic interaction terms are coupled by an explicit mass matrix  
\bea 
\label{lsg} 
\tilde V_{k}(\underline{\phi}) =  
\hf \underline{\phi}^{\rm T} \, {\underline{\underline{\tilde M}}}_k^2   
\underline{\phi} + {\tilde u}(k) \sum_{n=1}^{N_{L}} \cos (\beta \phi_n)  
\eea 
where $N_{L}$ is the number of the coupled fields  
(i.e. number of ``layers'') with the $O(N_{L})$ multiplet  
$\underline{\phi}=\left(\phi_{1}, \dots, \phi_{N_{L}}\right)$.  
The mass-matrix describes the interaction between the fields  
and is chosen here to be of the form 
\bea 
\label{mass_matrix_lsg} 
\hf \underline{\phi}^{\rm T} \,  
{\underline{\underline{\tilde M}}}_k^2 \underline{\phi} 
= \hf \tilde G_k \left(\sum_{n=1}^{N_{L}} \phi_n \right)^2 \,, 
\eea 
where $\tilde G_k$ is the strength of the inter-field interactions.

The WH--RG equation in LPA for the multi-component LSG model presented 
in our previous publications \cite{NaSa2006,Na2006,JeNaZJ2006,Na2008}
reads as 
\begin{equation} 
\label{WHdimlsg} 
(2+k \, \partial_k) \,\, \tilde V_k ({\underline\phi}) =  
- \frac{1}{4\pi} \ln \left[ {\mr{det}} \left(\delta_{ij}  
+ \tilde V_k^{ij}({\underline\phi}) \right) \right], 
\end{equation} 
where $\tilde V_k^{ij}({\underline\phi})$ denotes 
the second derivatives of the potential with respect to $\phi_i$, 
$\phi_j$. Inserting the ansatz \eq{lsg} into the RG equation  
\eq{WHdimlsg} the right-hand side becomes periodic, while  
the left-hand side contains both periodic and non-periodic  
parts. The non-periodic part contains only mass terms, so that  
we obtain a trivial tree-level RG flow equation for the  
dimensionless mass matrix  
\begin{equation} 
\label{mass_matrix_rg} 
\left(2 + k\partial_k \right) 
{\underline {\underline {\tilde M}}}^{2}_k = 0,  
\end{equation} 
which provides the trivial scaling $\tilde G_k = k^{-2} G$,  
where the dimensionful inter-field coupling $G$ remains constant  
during the blocking. We recall that in LPA there is no  
wave-function renormalization, thus the parameter $\beta$  
also remains constant during the blocking. Although the  
solution of \eqn{WHdimlsg} can only be obtained numerically,  
however, analytical results are also available using an  
approximation of \eqn{WHdimlsg} similarly to the MSG model. 
This is achieved by linearizing the WH--RG equation in the 
periodic piece of the blocked potential (not in the full potential),  
\begin{equation} 
\label{uv_wh_lsg} 
(2 + k \, \partial_k) {\tilde U}_k(\phi_1,..., \phi_{N_{L}}) 
\approx -  \frac{1}{4\pi} \, \,   
\frac{F_1(\tilde U_k)}{C}, 
\end{equation} 
where ${\tilde U}_k(\phi_1,..., \phi_{N_{L}}) = {\tilde u}(k) 
\sum_{n=1}^{N_{L}} \cos(\beta \, \phi_n)$ and  
$C$ and $F_1(\tilde U_k)$ stand for the constant and linear  
pieces of the determinant  
\begin{equation} 
\det[\delta_{ij} + {\tilde V}^{ij}_k]  
\approx C + F_1(\tilde U_k) + {\cal O}(\tilde U_k^2). 
\end{equation} 
The mass-corrected linearized WH--RG for the coupled periodic  
model \eq{lsg} reads as  
\begin{equation} 
(2+ k\partial_k ){\tilde u}(k) = \frac{\beta^2}{4\pi}  
{\tilde u}(k) 
\frac{k^2 + (N_{L}-1) G}{k^2 + N_{L} G} 
\end{equation} 
where $G$ and $\beta^2$ are scale-independent parameters  
and the solution can be obtained analytically 
\begin{equation} 
{\tilde u}(k) =  {\tilde u}(\Lambda)  
\left(\frac{k}{\Lambda}\right)^{\frac{(N_{L}-1)\beta^2}{N_{L} 4\pi}-2} 
\left(\frac{k^2 + N_{L} G} 
{\Lambda^2 + N_{L} G}\right)^{\frac{\beta^2}{N_{L} 8\pi}} 
\end{equation} 
where ${\tilde u}(\Lambda)$ is the initial value for the 
Fourier amplitude at the UV cutoff $\Lambda$. The critical  
frequency which separates the two phases of the model can be  
read out directly as   
\begin{equation} 
\label{laydep_m} 
\beta^2_{c}(N_{L}) = \frac{8\pi N_{L}}{N_{L}-1}. 
\end{equation} 
For $N_{L}=1$ the coupled model \eq{lsg} reduces to the massive 2D SG 
model with $\beta^2_c = \infty$ which indicates the absence of
the KTB phase transition. For $N_{L}\to\infty$ the coupled SG model 
behaves like a massless 2D-SG model with the critical frequency 
$\beta_c^2 = 8\pi$. Our goal is to consider whether the determination
of the critical frequency \eq{laydep_m} does depend on the choice of
renormalization scheme.
 
Since the P--RG method fails to determine the correct IR behavior 
of the one-component MSG model, here we do not apply the 
P--RG method to study the properties of the multi-component 
LSG model. Therefore, let us turn directly to the RG analysis of the 
LSG model \eq{lsg} by the  EAA--RG method with the optimized 
regulator. Then the RG flow equation for the local potential
of the $N_{L}$-component scalar field reads as
\beq 
\label{litim_rg_lsg} 
(2+k \partial_k) {\tilde V}_k =  \frac{1}{4\pi}  
\mathrm{Tr} \left( 
\frac{\delta_{ij}}{\delta_{ij}+\tilde V^{ij}_k} 
\right), 
\eeq 
where $\tilde V^{ij}_k =\partial_{\phi_i}\partial_{\phi_j} 
{\tilde V}_k$. Using the same machinery as in case of the  
WH--RG, the mass-corrected linearized form of the optimized  
regulator RG for the LSG model \eq{lsg} reads as
\begin{equation} 
(2+ k\partial_k ){\tilde u} = \frac{\beta^2}{4\pi}  
{\tilde u} 
\frac{k^4 + 2(N_{L}-1)G k^2 + N_{L}(N_{L}-1) G^2}{(k^2 + N_{L} G)^2}, 
\end{equation} 
where $G$ and $\beta^2$ are scale-independent parameters  
and the solution can be obtained analytically, 
\begin{eqnarray} 
{\tilde u}(k) &=&  {\tilde u}(\Lambda)  
\left(\frac{k}{\Lambda}\right)^{\frac{(N_{L}-1)\beta^2}{N_{L} 4\pi}-2} 
\left(\frac{k^2 + N_{L} G} 
{\Lambda^2 + N_{L} G}\right)^{\frac{\beta^2}{N_{L} 8\pi}} ,
\nonumber \\ 
&&\exp\left( \frac{\beta^2}{8\pi}\left[\frac{G}{k^2 + N_{L} G}-  
\frac{G}{\Lambda^2 + N_{L} G}\right]\right) ,
\end{eqnarray} 
where ${\tilde u}(\Lambda)$ is the initial value for the 
Fourier amplitude at the UV cutoff $\Lambda$. This gives 
the same $N_{L}$-dependent critical frequency \eq{laydep_m} 
as that  obtained by the WH--RG method.  
 
In general, any RG method which is applicable to predict the IR 
behavior for the one-component massless and massive SG models,
like the WH--RG, the EAA--RG with either the optimized or the quartic  
regulators, and the functional CS--RG methods, is assumed to produce 
the same $N_{L}$-dependent critical frequency \eq{laydep_m} for the 
LSG model \eq{lsg}. This can be understood by using a suitable 
$O(N_{L})$ rotated form of the coupled LSG model where the mass matrix 
is diagonal. Since the rotation leaves the phase structure unchanged, 
therefore, instead of the original model one can investigate the 
rotated one where the single massive 2D--SG field can be considered 
perturbatively which results in an effective SG-type model. 
This strategy can be applied for the multi-component coupled LSG 
model with arbitrary number of components ($N_L >1$), hence, in 
the lowest order of the perturbative treatment the corresponding 
effective theory is always an SG-type model and, consequently, the
various linearized RG equations produce the same critical frequency.

\section{Summary} 
\label{sum}
The renormalization of sine--Gordon (SG) type periodic scalar field 
models with explicit mass terms, i.e., the massive sine--Gordon (MSG)
and the multi-component layered sine--Gordon (LSG) models have been 
investigated by various functional renormalization group (RG) methods 
using the local potential approximation (LPA). Our aim was to compare the 
Wegner--Houghton, the Polchinski, the functional Callan-Symanzik methods 
and the effective average action RG method with various regulator 
functions and to investigate the scheme-dependence of the low-energy 
behavior of SG type models. In particular, our goal was to consider 
under which conditions is it  possible to determine scheme-independently 
three physical parameters, the critical frequency $\beta^2_c$ of the SG 
theory, the critical ratio $u/M^2$ of the MSG model and the layer-dependent 
critical value $\beta^2_c(N_L)$ of the LSG model.

Even if the effects of the truncation of the functional subspace 
in which the effective potential is sought for are under control, the 
RG flow depends on the particular choice of the renormalization scheme 
due to the different ways the quantum fluctuations are eliminated. 
Nevertheless no scheme-dependence is expected if the UV scaling laws or 
the (exact) IR physics are considered. UV scaling laws are invoked from 
the linearized flow equations being insensitive to the choice of the 
renormalization scheme. The IR physics is obtained by integrating out 
all the quantum fluctuations and the result does not depend on the 
manner how this happened if no approximations (e.g. a truncation of 
the functional subspace) were used. 
 
We demonstrated that the RG flows of the Wegner--Houghton, the  
functional Callan-Symanzik and the effective average action RG methods 
are similar for polynomial models (in the phase with unbroken $Z_2$ 
symmetry), for the generalized SG model (in its both phases) as well 
as for the MSG model. Those methods enable one to determine the IR 
physics, the phase structure of the MSG model in a reliable manner, 
in good agreement with density matrix and lattice results. It was also 
shown that the Polchinski RG method is inappropriate to determine the
RG flow beyond the mass scale in the deep IR limit. While this is not 
a drawback in case of polynomial models and the generalized SG model 
(with vanishing mass) it disables one to determine the phase structure 
of the MSG model quantitatively. 
 
The phase structure of the MSG model (which is the Bose form of QED$_2$) 
obtained by classical and quantum analysis were also compared. It is 
known that QED$_2$ has two phases, the strong $g>>m$ and the weak $g<<m$ 
coupling phase, where $m$ is the fermion mass and $g$ is the coupling 
between the gauge and the fermion field. The critical ratio $[m/g]_c$ 
which separates the phases of QED$_2$ obtained at the classical level 
was found to be two times smaller than that determined by RG methods. 
Above (below) the critical ratio, the periodicity of the bosonized model 
(i.e. for the MSG model) is broken spontaneously (explicitly). Therefore, 
the region of spontaneous symmetry breaking in the parameter space is 
reduced at quantum level as compared to the classical one which is in 
agreement with the general assumption, that the effect of quantum 
fluctuations always shrinks the region of spontaneous symmetry breaking. 
 
On the one hand, we showed that the critical frequency $\beta^2_c$ 
which separates the phases of an SG-type model which undergoes a 
Kosterlitz-Thouless-Berezinskii (KTB) type phase transition similar to 
the two-dimensional SG model can be obtained exactly (and scheme 
independently) by the RG flow linearized around the UV Gaussian fixed 
point. This is the consequence of the extension of the UV scaling 
region down to the vicinity of the crossover region separating the UV 
and IR scaling regions. As examples we considered the renormalization 
of the one-component two-dimensional SG theory and the multi-component 
LSG model where the two-dimensional SG fields of the various layers 
are coupled by an explicit mass matrix \eq{lsg}. On the other 
hand, it was also shown that the linearized RG equations produce a 
scheme-dependent critical ratio $[u/M^2]_c$ for the MSG model 
(i.e., $[m/g]_c$ for QED$_2$) and only the ``exact'' RG (where LPA 
was the only approximation used) gives the same (scheme-independent) 
result which coincides with the critical ratio determined by density 
matrix and lattice methods. Therefore, this demonstrates that a 
KTB-type phase transition of an SG-type model is a crossover between 
the UV and IR scaling regimes where the linearization around the 
Gaussian UV fixed point gives reliable results but the Ising-type 
phase transition of the MSG model is found to be an IR one where one 
has to solve the full RG equation in order to determine the exact 
transition point. 

Finally, let us conclude with two general results which are not 
restricted to the particular models investigated. We have shown
that in some cases it is possible to extend the validity of the 
UV linearized RG flow down to the vicinity of the crossover region 
separating the UV and IR scales which can produces physical 
parameters (such as the critical frequency $\beta^2_c$ of the SG 
and the layer number dependent critical value $\beta^2_c(N_L)$ of 
the LSG models) independently of the particular choice of the 
renormalization scheme. This is opposed to the general assumption 
namely that once approximations are used non-trivial fixed 
points and their critical behavior should be scheme-dependent. 
It was also shown that in case of spontaneous symmetry breaking 
the Maxwell construction represents a strong constraint on the RG
flow which results in a superuniversal IR behavior and as a 
consequence, scheme-independent IR results (such as the critical
ratio $[u/M^2]_c$ of the MSG model) can be obtained even
if the local potential approximation is used. This receives 
important application in any case where the low energy effective 
theory can only be determined by non-perturbative methods like 
functional RG approaches and a spontaneous symmetry breaking 
infuences the low energy behavior. Our results indicate that
scheme-independent results can be obtained even in the LPA.

\section*{Acknowledgment} 
Fruitful discussions with G. Delfino are warmly acknowledged. We also thank 
G. Mussardo and P. Sodano for several discussions. A.T. has been supported by 
the ESF grant INSTANS and by the MIUR projects ``Quantum Field Theory and 
Statistical Mechanics in Low Dimensions'' and ``Quantum Noise in Mesoscopic 
Systems''.

\appendix
 
\section{Renormalization Group Methods for $d=2$ in the Local Potential
 Approximation } 
\label{rgm}
In this appendix we settle our notations and remind the reader on some 
well-known features of the frequently used renormalization schemes in 
the local potential approximation (LPA) for two-dimensional Euclidean 
one-component scalar field theories. In principle the various 
renormalization schemes are constructed in such a manner that the RG 
flow  starts at the bare action and provides the effective action in 
the IR limit, so that the physical predictions (e.g. the critical 
exponents) are independent of the renormalization scheme particularly 
used. Nevertheless, scheme-dependence may appear in the RG flow at 
intermediate scales due to the different manners the quantum fluctuations
are eliminated in the various RG approaches. But even the physical 
predictions at the IR scales may depend on the used renormalization 
scheme if additional approximations are involved like the improperly
strong reduction of the functional subspace in which the local potential 
is sought for or the linearization of the RG flow equations at the 
Gaussian fixed point. Therefore, it is of relevance to clarify how far 
the results obtained are independent of the particular choice of the 
renormalization scheme used.

The use of the sharp cutoff, i.e. the Wegner--Houghton 
(WH) RG approach \cite{WeHo1973} makes the blocking transformation 
transparent and simple since the modes to be eliminated are well-defined.
The price of this clarity is the 
incompatibility with the gradient expansion. One possible solution 
for this problem could be the usage of the smooth momentum cutoff, where 
the higher frequency modes of the field are suppressed partially, but 
not eliminated. This can be realized by Polchinski's \cite{Po1984} 
method (P--RG) for the bare action and by the effective average action 
(EAA) RG approach \cite{RiWe1990,We1993,Mo1994,Li2001} with various 
types of regulator functions. As a rule, the solution of the truncated 
RG flow depends on the particular choice of the regulator. Another 
treatment of the RG which handles the effective action obtained by a 
suitable Legendre-transformation is represented by the functional 
Callan-Symanzik (CS--RG) or also called the internal space RG approach 
\cite{internal}, where the quantum fluctuations are separated according 
to their amplitudes instead of their frequencies or length scales. In 
this case the blocking procedure is performed in the space of the field 
variable and not in the space-time, which is understood as the external 
space.  
 
One of the advantages of the Wilsonian RG method is that the condensates 
which are generated in the RG flow can be treated in a simple manner. 
In fact, the condensates appear as non-trivial saddle points in the bare 
functional integral which can easily be detected and handled by expanding 
around the maximum of the integrand \cite{tree}. On the contrary, one 
always finds a convex effective action if the effects of the possible 
non-trivial saddle points are correctly incorporated during the 
Legendre-transformation, consequently, the Maxwell construction hides a 
large part of the dynamics generated by them. However, the RG methods 
based on the effective average action which interpolates between the bare 
action and the full quantum effective action provide generally an RG flow 
exhibiting singularity in a truncated functional subspace. The infrared 
(IR) singularity of the functional RG equation is supposed to be related 
to the convexity of the effective action for theories within a phase of 
spontaneous symmetry breaking \cite{We1993}.

\subsection{Wegner--Houghton RG} 
The blocking in momentum space, i.e. the integration over the field 
fluctuations with momenta of the magnitude between the UV scale $\Lambda$ 
and zero is performed in successive blocking  steps over infinitesimal 
momentum intervals $k\to k-\Delta k$ each of which consists of the splitting 
the field variable, $\phi=\varphi+\phi'$ in such a manner that $\varphi$ and 
$\phi'$ contain Fourier modes with $|p|<k-\Delta k$ and $k-\Delta k<|p|<k$, 
respectively and the integration over $\phi'$ leads to the Wegner--Houghton
(WH) RG equation 
\cite{WeHo1973}  
\beq 
\label{WHdim} 
\left(2 + k\partial_k \right) {\tilde V}_k ( \phi) =  
-\frac1{4\pi}\ln\left(1 + {\tilde V}''_k (\phi) \right) 
\eeq 
with $\tilde V''_k(\phi) = \partial^2_{\phi} \tilde V_k(\phi)$ for the 
dimensionless local potential ${\tilde V}_k = k^{-2} V_k$ for $d=2$ 
dimensions in the leading order of the derivative expansion, in the LPA 
when $\phi$ reduces to a constant. (Below we suppress the notation of 
the field-dependence of the local potential.) The differentiation with 
respect to the field variable and the multiplication with $1+\tV_k''$ 
leads to the derivative form of the WH--RG equation, 
\beq\label{derwh} 
(2+k\partial_k)\tilde V'_k =  
-\tilde V''_k(2+k\partial_k)\tilde V'_k-\frac1{4\pi}\tilde V'''_k. 
\eeq 
This equation is obtained by assuming the absence of instabilities for 
the modes around the gliding cutoff $k$. The WH-RG scheme which uses the 
sharp gliding  cutoff $k$ can also account for the spinodal instability, 
which appears when the restoring force acting on the field fluctuations 
to be eliminated vanishes, $1+\tilde V''_k(\phi)=0$ at some finite scale 
$k_{\mr{SI}}$  and the resulting condensate generates tree-level contributions 
to the evolution equation. The saddle point $\phi'_0$ for the single  
blocking step $k\to k-\Delta k$ is obtained by minimizing the action,  
$S_{k-\Delta k}[\phi]=\min_{\phi'_0}\left(S_k[\phi + \phi_0']\right)$. 
The restriction of the space of saddle-point configurations to that of 
the plane waves  $\phi'_0 = \rho \cos(k_1 x)$  gives \cite{tree}
\begin{align} 
\label{treedim} 
{\tilde V}_{k-\Delta k}(\phi) = \min_\rho \left[\rho^2 +\hf 
\int_{-1}^1 du {\tilde V}_k(\phi + 2\rho \cos(\pi u)) \right] 
\end{align} 
in LPA, where the minimum is sought for the amplitude $\rho$ only. It 
was shown that the tree-level RG equation \eq{treedim} leads to the local  
potential 
\begin{equation}
\label{treewhpot}
\tilde V_{k} = -\hf \phi^2
\end{equation}
which can also be obtained as the solution of 
\begin{equation}
\label{sieq}
1+\tilde V''_k(\phi)=0.
\end{equation}
If SI occurs during the RG flow at some scale $k_{\mr{SI}}>0$, then
Eqs. \eq{WHdim}  and \eq{derwh} can be applied only for scales 
$k>k_{\mr{SI}}$, and the tree-level renormalization should be performed 
at scales $k<k_{\mr{SI}}$. The right hand side of Eq. \eq{WHdim}  
develops a singularity when the SI occurs, but Eq. \eq{derwh} does 
not develop such a singularity and its solution mathematically  
extends to $k\to 0$. It is interesting to notice, that \eqn{derwh} 
yields  the fixed-point equation
\begin{equation}
\label{fpwh}
2\tV_* +[ \tV_*^\prime]^2 + \frac{1}{4\pi} \tV_*^{\prime\prime}=c_1
\end{equation}
with the arbitrary constant $c_1$, exhibiting the trivial solution 
$\tV_*=c_1/2$ (Gaussian fixed point) and  
\begin{equation}
\label{whfppot}
\tV_* = - \hf \phi^2
+ \frac{1}{8\pi}+\hf c_1
\end{equation} 
which is equivalent to the fixed point potential \eq{treewhpot}.

\subsection{Polchinski's RG} 
In Polchinski's RG (P--RG) method \cite{Po1984} the realization of the 
differential RG transformations is based on a non-linear generalization of 
the blocking procedure using a smooth momentum cutoff. In the infinitesimal 
blocking step the field variable $\phi$ is split again into the sum of a 
slowly oscillating IR and a fast oscillating UV components, but both fields  
contain now low- and high-frequency modes, as well, due to the smoothness of 
the cutoff. Above the moving momentum scale $k$ the propagator for the IR 
component is suppressed  by a properly chosen smooth regulator function $K(y)$ 
with $y=p^2/k^2$, $K(y)\to 0$ if $y>>1$, and $K(y)\to 1$ if $y<<1$. The P--RG 
equation in LPA for $d=2$ dimensions reads as 
\beq\label{polch} 
(2+k\partial_k)\tilde V_k = -[\tilde V'_k]^2 K'_0+\tilde V''_k I_2, 
\eeq 
where $K'=\partial_y K(y)$, $K'_0 = \partial_{y} K(y)\vert_{y=0}$ 
and $I_2= (1/4\pi)\int_0^\infty K'(y)$. The parameters $K'_0$ and $I_2$ 
can be eliminated by the rescaling of the potential and the field variable, 
consequently, they do not influence the physics. In order to make the 
comparison of the RG flows obtained by various RG methods straightforward, 
we choose $I_2 \equiv -\frac1{4\pi}$ and $K_0^\prime=-1$ for which the 
linearized forms  of \eqn{WHdim} and \eqn{polch} and the UV scaling laws
obtained by WH--RG and P--RG are identical.
Then the differentiation of both sides of \eqn{polch} with respect to 
the field variable $\phi$ yields 
\beq\label{derpolch2} 
(2+k\partial_k)\tilde V'_k =  
2\tilde V''_k\tilde V'_k-\frac1{4\pi}\tilde V'''_k 
\eeq 
being independent of the regulator function $K(y)$ and differing of the 
WH--RG equation \eq{derwh} by the term $-\tilde V''_k k\partial_k\tilde V'_k$
with opposite sign for the non-linear term. 
 
Let us note on the one hand, that the P--RG method treats all quantum 
fluctuations below and above the scale $k$ on the same footing. Therefore, 
even if there occurs a scale $k_{\mr{SI}}$ at which $1+\tV_k''$ exhibits 
zeros, one cannot decide unambiguously at what scale should one turn to 
tree-level renormalization. On the other hand, \eqn{polch} with the choice 
of the parameters $K_0^\prime=-1$, $I_2=-\frac{1}{4\pi}$ leads to the 
fixed-point equation similar to that of \eq{fpwh} with $c_1=0$ 
\begin{equation}
\label{fppolch}
2\tV_* -[ \tV_*^\prime]^2 + \frac{1}{4\pi} \tV_*^{\prime\prime}=0
\end{equation}
and exhibits trivial fixed-point solutions: the Gaussian one 
($\tV_* = \mr{const}$) and the high-temperature (or infinitely massive) 
fixed point
\begin{equation}
\label{polchfppot}
\tV_* = + \hf \phi^2 - \frac{1}{8\pi}
\end{equation} 
which is only accounted for the P--RG method. Let us note, that 
\eq{polchfppot} is similar to \eq{whfppot} which is obtained in the 
framework of the WH--RG method but with opposite sign. For dimensions 
$d=2$ it was shown in \cite{Mo1995} that non-trivial fixed point 
solutions in LPA are either singular at finite $\varphi$ or periodic.

\subsection{Effective average action RG}\label{eaarg} 
The effective average action (EAA) RG method \cite{RiWe1990,We1993,Mo1994} 
has grown out of the idea of coarse-graining the quantum fields and it 
interpolates between the bare action and the full quantum effective action. 
The scale-dependent effective average action $\Gamma_k$ satisfies the 
functional differential equation
\beq\label{eff_rg_func} 
k \partial_k \Gamma_k = \hf \mathrm{Tr}  
\left( \Gamma^{(2)}_k + R_k \right)^{-1}  
k \partial_k R_k 
\eeq 
where $\Gamma^{(2)}_k$ denotes the second functional derivative of 
the effective action. Here  $R_k$ is a properly chosen IR regulator 
function which fulfills a few basic constraints to ensure that $\Gamma_k$ 
approaches the bare action in the UV limit ($k\to\Lambda$) and the full 
quantum effective action in the IR limit ($k\to 0$) and to guarantee 
that no IR divergences are encountered in the presence of massless 
modes. In the present paper we shall use $R_k(p^2)\equiv p^2 r(y)$ 
with the power-law regulator $r(y)=y^{-b}$ $(b>1)$ \cite{Mo1994} 
and the optimized regulator \cite{Li2001} 
$r(y) = \left(\frac{1}{y} -1 \right) \Theta(1-y)$ ($\Theta$ denotes 
the Heaviside step-function) where $y=p^2/k^2$. For $d=2$ dimensions 
Eq. \eq{eff_rg_func} can be rewritten in the LPA as  
\beq
\label{general_effective_rg} 
(2+k \partial_k) {\tilde V}_k = - \frac{1}{4\pi}  
\int_{0}^{\Lambda^2/k^2} \mathrm{d}y 
\frac{r' \, y^2}{y(1+r) +{\tilde V''}_k} 
\eeq 
with $r' = dr/dy$ for the dimensionless local potential.
The truncation of the basis set of functions in which the local potential 
is expanded may introduce a dependence of  the physical results (e.g. the
critical exponents) on the particular choice of the regulator function.
In the literature several optimization procedures have been proposed
in order to achieve better convergence of the critical exponents with the  
removal of that truncation e.g., Refs. \cite{LiPoSt2000,Li2001}. For 
example, if one compares various RG schemes, it has been argued that the 
fastest convergence can be achieved by using the optimized regulator 
\cite{Li2001}. Concerning the power-law regulator it was argued that 
the optimal choice of the parameter $b$ is met when the minimum value 
\begin{equation}
\label{cb}
C(b)= \frac{b}{(b-1)^{(b-1)/b}}
\end{equation}
is maximal, which happens for $b=2$, i.e., for the quartic regulator. For 
arbitrary parameter value $b$, the propagator in the right hand side of 
Eq. \eq{general_effective_rg} may develop a pole at some scale $k_{\mr{SI}}$ 
and at some value of the field $\phi$  for which $\tV_k''(\phi)=-C(b)$ 
holds, which signals the occurring of SI. It was shown that in such a case 
one has to seek the local potential for $k<k_{\mr{SI}}$ by minimizing 
$\Gamma_k$ in the subspace of inhomogeneous (soliton like) field 
configurations and ends up with the result \cite{We1993} 
\begin{equation}
\label{eaafppot}
\tV_k=-\hf C(b)\phi^2
\end{equation}
of parabolic shape, the solution of the equation 
\begin{equation}
\label{eaasi}
C(b)+\tV_k''(\phi)=0.
\end{equation}

For the quartic regulator the integral over $y$ can be performed and
Eq. \eq{general_effective_rg} rewritten in the explicit form
\begin{align}  
\label{morris_b2_rg}
&(2+k \partial_k) {\tilde V}_k 
= - \frac{1}{2\pi\sqrt{|1-(\hf\tilde V''_k)^2|}} \times 
\\[2ex]  
&\times \left\{ \begin{array}{cc}   
\mathrm{arctg} 
\left(\frac{\hf\tilde V''_k}{\sqrt{1-(\hf\tilde V''_k)^2}}\right)     
- \frac{\pi}{2}, 
&   
{\rm{for~}} \vert(\hf\tilde V''_k)\vert < 1 \\[2ex]  
\frac{1}{2} 
\ln \left(\frac{\hf\tilde V''_k - \sqrt{(\hf\tilde V''_k)^2 -1}} 
{\hf\tilde V''_k + \sqrt{(\hf\tilde V''_k)^2 -1}}\right),                 
&  
 {\rm{for~}} (\hf\tilde V''_k) > 1  
\end{array} \right. \nonumber  
\end{align}  
when the limit $\Lambda \to \infty$ is taken. Making use of l'Hospital's 
rule it is straightforward to show that the right hand side of 
\eq{morris_b2_rg} tends to $\frac{1}{4\pi}$ for $\hf\tilde V''_k \to 1$ 
from either below or above and, consequently, the couplings behave 
non-analytically at such scale $k_+$, but do not diverge to infinity.  
However, if there exists a scale $k_{-}>0$ for which $\hf\tilde V''_k\to -1$, 
then the right hand side of \eq{morris_b2_rg} becomes infinite which 
indicates the appearance of SI due to vanishing of the inverse of the IR 
regulated propagator.
Therefore the scale $k_-$ can be identified with the scale of the SI, 
$k_- \equiv k_{\rm{SI}}$. The scale $k_+$ may exist even if no SI arises 
during the flow, it is an artifact of the regulator used, but of no 
physical significance. We shall demonstrate this latter case for the 
one-component scalar field theory with polynomial interaction. The 
two-dimensional generalized sine-Gordon model shall demonstrate another 
interesting case when the scales $k_\pm$ are equal. 

It is worthwhile noticing that Eq. \eq{general_effective_rg} with the 
power-law regulator leads to the WH-RG equation \eq{WHdim}, and 
\eqn{eaasi}  leads to \eqn{sieq} for $b=1$ as well as for $b\to \infty$
in the limit $\Lambda\to \infty$. This feature of the case $b=1$ holds 
only for $d=2$.

The RG equation with the optimized regulator in the LPA reads as  
\beq\label{litim_rg} 
(2+k \partial_k) {\tilde V}_k = - \frac{1}{4\pi}  
\frac{\tilde V''_k}{1+\tilde V''_k} 
\eeq 
where a field-independent constant has been added to the right hand 
side of the equation. Let us note that if SI occurs, the right hand 
side of \eqn{litim_rg} becomes infinite at the scale $k_{\mr{SI}}$, 
so that all or some of the couplings should tend to infinity at this
scale. One should again turn to the explicit treatment of SI. The 
equation obtained by deriving \eq{litim_rg} with respect to the field 
and multiplying its both sides by $(1+\tV_k'')^2$ leads to the 
fixed-point equation $2(1+\tV_*'')^2\tV'_* =  - \frac{1}{4\pi}\tV_*'''$, 
which exhibits the trivial solution $\tV_*=\mr{const}$ (Gaussian fixed 
point), and also the parabolic solution \eq{whfppot}. Let us note, it 
was demonstrated that Eqs. \eq{litim_rg} and \eq{polch} give the same 
critical exponents for the O(N) symmetric scalar theory in three
dimensions \cite{litimpolch} and the two RG equations can be 
transformed onto each other via a suitable Legendre transformation 
\cite{Mo2005}. However, the RG trajectories and the singularity 
structures of \eq{litim_rg} and \eq{polch} could be different.

\subsection{Functional Callan-Symanzik RG} 
In the functional Callan-Symanzik (CS--RG) type internal space 
RG method \cite{internal}, the successive elimination of the 
field fluctuations is performed in the space of the field variable 
(internal space) as opposed to the usual RG methods where the 
blocking transformations are realized in either the momentum or 
the real (external) space. This can be achieved by introducing an 
additional mass term into the bare action, 
\beq 
S_\lambda[\phi] = S_B[\phi]+\hf \lambda^2 \phi^2 
\eeq 
with the control parameter $\lambda$. For $\lambda=\lambda_0$  
being of the order of the UV cutoff the large-amplitude  
fluctuations are suppressed and decreasing the control 
parameter $\lambda$ towards zero, they are gradually  
accounted for. The functional evolution equation for the  
effective action is 
\beq\label{mdint} 
\lambda \partial_{\lambda}{\Gamma}_\lambda = 
\hf\mr{Tr} \left[\lambda^2 
+{\Gamma}^{(2)}_{\lambda}\right]^{-1} 2\lambda^2, 
\eeq 
where ${\Gamma}^{(2)}_{\lambda}= 
\delta^2{\Gamma}_\lambda/\delta\varphi\delta\varphi$.  
Eq. \eq{mdint} is equivalent to the RG equation \eq{eff_rg_func} 
for the EAA with the power-law regulator function with $b=1$ for 
$\lambda=k$. By using Eq.\eq{general_effective_rg} the functional 
Callan-Symanzik RG equation for the one-component scalar field 
theory for dimensions $d=2$ in the LPA reads 
\bea   
\label{internal_step1} 
(2+\lambda \, \partial_{\lambda}) \,\, \tilde V_{\lambda} =  
- \frac{1}{4\pi}  
\ln \left(\frac{1 + \tilde V''_{\lambda}} 
{\Lambda^2/\lambda^2 + 1 +\tilde V''_{\lambda}} \right), 
\eea  
where $\tilde V_{\lambda}\equiv\lambda^{-2} V_{\lambda}$ is the  
dimensionless scale-dependent effective potential. Adding the 
field-independent term $-1/(4\pi)\ln(\Lambda^2/\lambda^2)$ 
to the right hand side of Eq.\eq{internal_step1}, one can take 
the limit $\Lambda\to\infty$, 
\bea  
\label{internal_rg}  
&(2+\lambda \, \partial_{\lambda}) \,\, \tilde V_{\lambda} =  
- \frac{1}{4\pi}  
\ln \left(1 + \tilde V''_{\lambda} \right). 
\eea  
This equation is mathematically equivalent to the two-dimensional 
WH--RG equation in the LPA assuming the equivalence of the scales 
$\lambda \equiv k$. However, for dimensions $d\neq 2$ the functional
Callan-Symanzik RG and the WH--RG differ from each other. Assuming 
the above mentioned equivalence of the scales $\lambda$ and $k$, there 
occurs the same singularity in the right hand side of \eq{internal_rg} as 
the one in the WH-RG approach. Therefore, the functional Callan-Symanzik 
RG  signals the SI with the vanishing of the argument of the logarithm 
in the right hand side of \eq{internal_rg}. The solution of 
\eq{internal_rg} provides the scaling laws down to the scale $k_{\mr{SI}}$ 
and one has to turn to the tree-level renormalization with the help of 
the WH-RG approach in order to determine the IR scaling laws.

\section{Comparison of RG schemes for polynomial interaction}
\label{polyn} 
Here, we compare the applicability of the frequently used functional RG 
methods to two-dimensional one-component Euclidean scalar fields with 
polynomial self-interaction. We show that the flow of the polynomial 
model determined by the P--RG differs of that obtained by other RG methods 
in the deep IR regime. For the polynomial local potential
\bea  
\label{polynom}  
\tilde V_{k}(\phi) = \sum_{n=1}^N \, \frac{1}{(2n)!} \,  
{\tilde g}_{2n}(k) \,\, \phi^{2n} 
\eea  
the scale-dependence is encoded in the dimensionless coupling constants 
$\tilde g_{2n}(k)$, related to their dimensionful counterparts via 
$k^2 \tilde g_{2n} = g_{2n}$. The bare action exhibits the $Z_2$ 
symmetry $\phi\to -\phi$, and it is well-known that there exist two 
phases of such a model with either unbroken or broken $Z_2$ symmetry.

\begin{figure}[ht] 
\includegraphics[width=8cm]{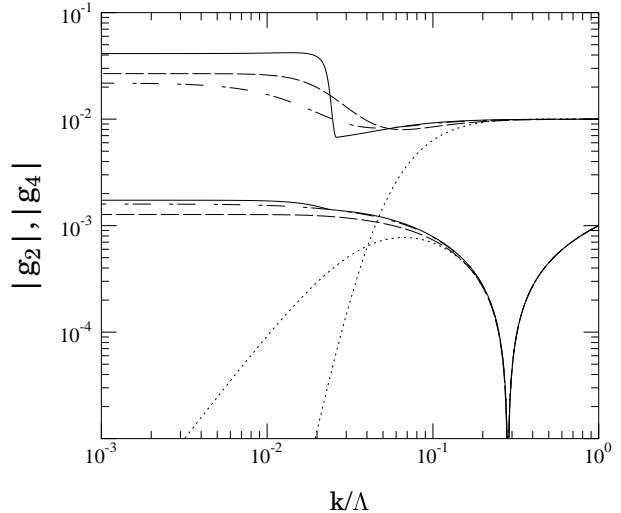} 
\caption{The RG flow of the dimensionful couplings $g_{2}(k)$  
and $g_{4}(k)$ obtained by various RG methods for the same  
initial condition $g_{2}(\Lambda)=-0.001$, $g_{4}(\Lambda)=0.01$,
$g_{2n>4}(\Lambda)=0$. The full, dashed, and dashed-dotted lines 
correspond to the EAA--RG with quartic regulator, the EAA--RG with 
optimized regulator, and the WH--RG flow, respectively and the dotted 
lines represent the flow obtained by the P--RG method. The mass $g_2(k)$ 
changes its sign in any cases at $(k/\Lambda) \sim 3 \cdot 10^{-1}$ 
for the given initial conditions. If the flow is determined by the 
quartic regulator RG, $g_4(k)$ is non-analytic at the momentum scale 
$k_{+} \approx 2.64 \cdot 10^{-2}$ for which $k_{+}^2 = g_2(k_+)$. 
\label{fig_poly}} 
\end{figure}

Let us start with the discussion of the RG flow in the phase with 
unbroken $Z_2$ symmetry. Inserting the potential \eq{polynom} into any 
of the RG equations \eq{derwh}, \eq{litim_rg}, \eq{morris_b2_rg}, 
\eq{derpolch2} and expanding both sides of them into Taylor-series in 
the dimensionless field-variable  $\phi$, one arrives at a coupled set 
of ordinary, non-linear differential equations for the couplings 
$\tg_{2n}(k)$. These have been solved numerically for the same initial  
condition $g_{2}(\Lambda)=-0.001$, $g_{4}(\Lambda)=0.01$, $g_{2n>4}=0$
at the UV cutoff $\Lambda$ which ensured that no SI occurred, i.e. the 
inequality $k^2 + g_{2}(k)>0$ has been kept during the flow. 
The RG flow of the couplings $g_2(k)$ and $g_4(k)$ for various RG 
methods and for $N=10$ is plotted in \fig{fig_poly}. There were no 
appreciable changes in the results when we increased $N$ further. 
The WH--RG, the EAA--RG with optimized regulator, and the EAA--RG with 
quartic regulator give qualitatively the same results, the dimensionful 
couplings become constant in the IR limit. Let us note that $g_4(k)$ is 
non-analytic at the scale for which $k_{+}^2= g_2(k_+)$ if the flow is 
determined by the quartic regulator RG. 

For the P--RG method the flow of the couplings in the IR region 
($k\to 0$) differs of that obtained by WH--RG and EAA--RG methods 
(see \fig{fig_poly}). The RG trajectories tend to the trivial 
high-temperature (or infinitely massive) fixed point \eq{polchfppot}
which is accounted only for the P--RG method. (Let us note, that in
three-dimensions the critical behavior is dominated by the non-trivial
Wilson-Fisher fixed point.) However, one may argue that the RG analysis 
is physically uninteresting in the IR limit since below the momentum 
scale $k^2 = g_2(k)$ the mass term suppresses the quantum fluctuations 
in the propagator and after a rather small transient domain, the RG 
flow obtained by the other RG methods becomes trivial, i.e. all the 
dimensionless couplings ($\tilde g_2$, $\tilde g_4$) scale as 
$\sim k^{-2}$ (for $d=2$), consequently, the corresponding dimensionful 
parameters ($g_2$, $g_4$) tend to constants, see \fig{fig_poly}. 
It is also demonstrated in \fig{fig_poly}, that even the P--RG 
shows up the change of the sign  of $g_2(k)$ because it occurs above 
the mass-scale. However, in this paper we show that there may be 
special situations such as for example the case of the MSG model where 
one has to go beyond the mass-scale in order to map out the phase 
structure of the model in a reliable manner. In such a case the P--RG 
method is inappropriate for quantitative analysis.

The discussion above has been restricted to the phase with unbroken 
$Z_2$ symmetry, when no SI occurs during the RG flow. The RG flow has
also been investigated in the literature for the phase with spontaneously 
broken $Z_2$ symmetry  in the frameworks of both the WH--RG involving 
tree-level renormalization and the EAA--RG. The truncated WH-RG flow becomes 
numerically unstable, i.e. the couplings $\tg_{2n}(k)$ start to heavily 
oscillate approaching the scale $k_{\mr{SI}}$ from above where the SI occurs. 
But it has been shown that the tree-level evolution for RG trajectories 
started below the scale $k_{\mr{SI}}$ run into the universal IR fixed-point 
potential \eq{treewhpot}. Applying the explicit treatment of SI proposed 
in \cite{We1993}, one would obtain similar results by the EAA-RG. It has 
been argued in \cite{We1993} that the dimensionless IR effective potential 
for $k<k_{\mr{SI}}$ reads as the fixed-point potential \eq{eaafppot} (c.f. 
Appendix \ref{eaarg}). It was also demonstrated in \cite{TeWe1992} that 
the appearance of SI can be avoided by a suitable rescaling  of the RG 
equations at least in case of the polynomial scalar theory. However, an 
attractive IR fixed point appears in the rescaled RG flow which can be 
identified as the effective potential \eq{eaafppot}.


\end{document}